\documentstyle[preprint,eqsecnum,prb,aps,tighten]{revtex}
\begin{document}
\preprint{UBCTP-93-026}
\draft
\title{Equal Time Correlations in Haldane Gap Antiferromagnets}
\author{Erik S.\ S\o rensen$^a$ and Ian Affleck$^{a,b}$ }
\address{$^{(a)}$Department of Physics and $^{(b)}$Canadian Institute
for Advanced Research}
\address{
University of British Columbia, Vancouver, BC, V6T 1Z1, Canada}
\date{\today}
\maketitle
\begin{abstract}
The $S=1$
antiferromagnetic Heisenberg chain both with and without single ion 
anisotropy is studied.
Using the recently proposed density matrix renormalization group
technique we calculate the energy gaps
as well as several different correlation functions.
The two gaps, $\Delta_{||}, \Delta_\perp$, along with associated
correlation lengths and velocities are determined.
The numerical results are shown to be in 
good agreement with theoretical predictions
derived from the nonlinear sigma model and a free boson model.
We also study the $S=1/2$ excitations that occur at the ends
of open chains; in particular we study the behavior associated with
open boundary conditions, using a model of $S=1/2$ spins
coupled to the free bosons.
\end{abstract}
\pacs{75.10.-b, 75.10.Jm, 75.40.Mg}

\section{Introduction}
Following the initial conjecture of Haldane~\cite{haldane}
it is by now well established both experimentally~\cite{expgap,renard87} and
theoretically~\cite{numgap,tak1,sakai2,white2,tak93,jphysc}
that integer spin antiferromagnetic chains have a gap 
as opposed to chains with half-integer spin.
The spectrum is that of a disordered singlet ground-state
with a gap, $\Delta$, to a triplet excitation. 
The disordered Haldane phase is known to have a hidden
order parameter, the so called string order parameter~\cite{nijs89}.
One of the best candidates for such a Haldane gap system is
Ni(C$_2$H$_8$N$_2$)$_2$NO$_2$(ClO$_4$) (NENP).
The ratio of inter- to intra chain coupling in
this compound is estimated~\cite{renard2}
to be less than $4\times10^{-4}$
and no ordering has so far been observed at any accessible
temperatures. 
NENP is not a completely isotropic compound. As revealed
by neutron scattering experiments the magnon spectrum
around wave vector $\pi$ is split into three
branches~\cite{expgap,regnault}
due to the effects of anisotropies. Electron spin resonance
studies~\cite{brunel,palme} have observed transitions between
the different components of the triplet confirming the expected
nature of the spectrum. Thus, the isotropic
Haldane gap, $\Delta$, is effectively split in three.
Single ion anisotropy, $D(S^z_i)^2$, splits the triplet
into two modes for fluctuations
perpendicular and parallel to the chain, respectively. In addition,
in-plane anisotropy, of the form
$E((S^x_i)^2-(S^y_i)^2)$, splits
the perpendicular mode by a small amount.
At present the best estimates of the gaps in 
NENP~\cite{regnault,brunel,palme}
are 2.50, 1.25, and
1.05 meV, respectively.
The dispersion relation and dynamical structure factor have also
been measured for NENP~\cite{renard87,renard88,renardC8,ma}.
The largest of the anisotropies, and the only one 
we shall be concerned with here, is the
single-ion anisotropy, $D$.
A simple Hamiltonian capturing the essential physics including
the single ion anisotropy is then
\begin{equation}
H=J\sum_i\{{\bf S}_{i}\cdot{\bf S}_{i+1}+D(S_i^z)^2\}.
\label{eq:h}
\end{equation}
As a function of $D$ the lowest lying gap will decrease. At a
critical value $D_c\sim 1$ the gap closes and a phase transition
between the disordered Haldane phase 
and a ``large-$D$" phase occurs~\cite{golinelli2}.
For a discussion of the phase diagram see Ref.~\onlinecite{nijs89}.
By fitting exact diagonalization results on chains of length up to
18 to inelastic neutron scattering results
Golinelli et al~\cite{golinelli1} have
determined the optimal values $J=3.75$ meV and $D/J=0.18$.
This value of $D$ is then clearly in the disordered Haldane phase.
Even including inter chain coupling, which changes the value of $D_c$,
$D/J=0.18$ remains in the disordered phase~\cite{sakai1}, consistent
with the fact that no ordering is observed in NENP at finite 
temperatures. Recently it has been shown that the Hamiltonian
Eq.~(\ref{eq:h}) should be corrected to account for the 
staggered structure of the NENP chains~\cite{partha93},
a fact that also leads to a staggered off-diagonal part
in the gyromagnetic tensor~\cite{gyro}. 

The longest chains one has studied by exact diagonalization
have 18 sites~\cite{golinelli2,golinelli3,tak93,haas}.
While it is possible to study somewhat longer chains
using quantum Monte Carlo (QMC) techniques~\cite{tak1,tak93}
the statistical errors
on these results will limit the detailed interpretation
of the data. Recently White~\cite{white1} has proposed
a density matrix renormalization group (DMRG) method that allows
one to study significantly longer chains with much higher accuracy
than what is obtainable using QMC methods.
At present  
it is not known how to work with the momentum
as a quantum number within the DMRG framework although
this would be very desirable and one is therefore restricted
to study quantities defined in real space. Another less important
restriction imposed by the DMRG is the fact that it works
notably better for chains with open ends. Recent progress~\cite{white3}
indicates that open ended systems can be tuned to mimic the
behavior of systems subject to periodic boundary conditions.
In this paper we calculate equal time correlation functions in
real space for the
$S=1$ antiferromagnetic chain for systems of up to 100 spins
using the DMRG method. The numerical results are then compared
to predictions obtained from the nonlinear sigma (NL$\sigma$) model
and a free boson model. Some of our results have previously been
presented elsewhere~\cite{us2}.

Equal time correlation functions have been calculated 
by QMC techniques~\cite{sogo,tak88,nomura,liang,meshkov1} and
by exact diagonalization~\cite{moreo,sakai2,sakai3,golinelli3}
for chains of length up to L=18. 
The lowest energy excitation as a function of momentum has
been calculated by QMC techniques~\cite{tak1,tak93} and
by exact diagonalization~\cite{golinelli2,golinelli3,tak93,haas}
for chains of length up to L=18.
The 
dynamic structure factor has been
studied by QMC techniques~\cite{deisz,meshkov1,meshkov2,deisz2}.
The dynamical structure factor, including anisotropy,
for both in plane and out of plane modes
has been studied by Golinelli et. al~\cite{golinelli3}
using exact diagonalization on chains with
$L\leq 16$, and by Haas et. al.~\cite{haas}
for the out of plane mode with $L\leq 18$.
While generally applicable the current state of the art
exact diagonalization studies are limited by the fact
that only very short chains can be studied, thus restricting
the accessible momentum values considerably. Secondly,
finite size effects are likely to be large. The finite size
effects can be improved upon by the application of extrapolation
techniques, (Shanks transformations), if enough data points
are available. Thus, values for the gaps can be extracted to a
very good precision. However, for the structure
factor, this technique can only be applied for a few
values of the momentum.
The QMC results, while applicable to considerably longer chains,
have statistical errors. 
The extraction of dynamical properties from QMC results requires
the analytical continuation of results obtained
at imaginary frequencies to real frequencies. This is typically done
using maximum entropy methods and is not a trivial matter.
While dynamical properties are not accessible yet using the DMRG method,
it is possible to obtain equal time correlations for very long chains
with an unprecedented accuracy. This huge gain in detail and precision
allows a detailed comparison to field theory and experiment not
previously possible.

We shall here only be concerned
with chains of even length subject to open boundary conditions.
All of our calculations are for chains described by the Hamiltonian
Eq.~(\ref{eq:h}) with $J/k_b$ =1 and $D/J=0.18$ or $D/J=0.0$.
For these chains
the ground-state is a singlet with even parity, $0^+$. Above the
ground-state is an exponentially low-lying triplet, $1^-$. In the
thermodynamic
limit the triplet and the singlet become degenerate and the
ground-state four-fold degenerate. The bulk of the calculations
of the correlation functions presented
here will be for a 100 site chain in
the $1^-$ state since this system conceptually is
the simplest. We have repeated the calculations for the
$0^+$ state and also performed the same calculations for
a 60 site system. In none of these cases were the correlation
functions seen to differ markedly from what we will describe below,
aside from boundary effects. In particular, all the
correlation lengths obtained were consistent.

We implement the DMRG
using density matrices of the size
$243\times243$ keeping 81 eigenvectors of these matrices at each
iteration.
First density matrices representing systems of different sizes
are computed using an infinite lattice method, then these
are combined to form a system of a fixed size, L=100, for
which all the correlation functions are calculated
using
a finite lattice method~\cite{white1}.
For a discussion of the numerical procedure we
refer the reader to Ref.~\onlinecite{white1,white4}.
The DMRG method for open chains leaves two good quantum numbers
the total $S^z$ component, $S^z_T$, and the parity, $P$, corresponding
to a reflection about the midpoint of the chain.
These are
conserved under iteration
and it is therefore possible to work within a
subspace defined by these two quantum numbers.
In the following
we characterize states solely by these two quantum numbers:
the total $S^z$ component, $S^z_T$, and the parity, $P$.
An additional symmetry, corresponding to a global spin flip
interchanging -1 and 1 while leaving 0 alone, introduces
another parity in the $S^z_T=0$ sector. Since most of our
calculations have been performed in the $S^z_T=1$ sector
we have not considered this symmetry.

In Section~\ref{sec:gl} we shall briefly review the free boson
model. Section~\ref{sec:gaps} is concerned with our results
for the energy gaps. In Section~\ref{sec:bulkcor}
we discuss the bulk correlation functions.
Section~\ref{sec:boundcor} addresses the effects that the
open ends have on the system and discusses the boundary
correlation functions.
Theoretical and numerical results are presented
for the two types of correlation functions in these two sections.
In Section~\ref{sec:struct} we discuss the equal time structure factor
in conjunction with experimental results and theoretical estimates.
Section~\ref{sec:sma} discusses the single mode approximation for this
problem.

\section{Free Boson Model}\label{sec:gl}
It is believed that the isotropic $S=1$ Heisenberg antiferromagnetic chain
can be approximately described, at low energies,
by the non-linear $\sigma$ (NL$\sigma$)
model~\cite{haldane}, with
Hamiltonian
\begin{equation}
H=\frac{v}{2}\int dx \left[ g {\bf l}^2+\frac{1}{g}
\left(\frac{\partial {\bbox{\phi}}}
{\partial x}\right)^2\right],\ \ g=\frac{2}{s},\ \ v=2Jas,
\label{eq:hnl}
\end{equation}
where $a$ is the lattice spacing.
Since we have set $\hbar=1$, the velocity, $v$, has dimension energy
times length.
The massive triplet of fields ${\bbox{\phi}}$ is restricted to have
unit magnitude, ${\bbox{\phi}}^2=1$, and ${\bbox{\phi}}$ and ${\bf l}$ describe
the sublattice and uniform magnetization respectively.
This Hamiltonian can be arrived at directly from a lattice
Hamiltonian~\cite{ian85}, by use of the relations
$\bbox{\phi}(2i+1/2)=[{\bf S}_{2i+1}-{\bf S}_{2i}]/(2s),\ \ 
{\bf l}(2i+1/2)=[{\bf S}_{2i}+{\bf S}_{2i+1}]/(2a)$.
We have then approximately ${\bf S_i}=(-1)^is{\bbox{\phi}}+a{\bf l}$, with
${\bf l}=1/(vg){\bbox{\phi}}\times(\partial {\bbox{\phi}}/\partial t)$.
The NL$\sigma$ model has Lagrangian density
\begin{equation}
{\cal L}=\frac{1}{2g}\left[\frac{1}{v}
\left(\frac{\partial {\bbox{\phi}}}{\partial t}\right)^2 
-v\left(\frac{\partial {\bbox{\phi}}}{\partial x}\right)^2\right].
\label{eq:lnl}
\end{equation}
It is possible to obtain a linear model in the spirit
of a phenomenological Landau-Ginzburg model describing the correct
physics at the
mean field level by introducing explicitly a mass term and a
quartic term for stability~\cite{ian} while lifting the non-linear
constriction ${\bbox{\phi}}^2=1$.
The $D$ term in Eq.~(\ref{eq:h}) describing
the single ion anisotropy will split the massive triplet into a
low-lying doublet, with gap $\Delta_\perp$, and a higher lying
singlet, with gap $\Delta_{||}$. One usually takes the $z$ axis
to be along the chain, describing the out of plane $||$ mode,
with the $x,y$ axis perpendicular to the chain describing
the in-plane $\perp$ mode. The phenomenological Lagrangian then becomes
\begin{equation}
{\cal L}=\sum_{i=1}^3\left[\frac{1}{2v_i}
\left(\frac{\partial\phi_i}{\partial t}\right)^2
-\frac{v_i}{2}\left(\frac{\partial\phi_i}{\partial x}\right)^2
-\frac{\Delta_i^2}{2v_i}(\phi_i)^2\right]-\alpha{\bbox{\phi}}^4.
\label{eq:fb}
\end{equation}
Here we have
$
{\bf S}_i\approx s\sqrt{g}(-1)^i{\bbox{\phi}}+a{\bf l},
$
with $g=g_\perp, g_{||}$ for the perpendicular and parallel modes,
respectively, and
\begin{equation}
{\bf l}=\frac{1}{v}{\bbox{\phi}}\times\frac{\partial{\bbox{\phi}}}{\partial t},
\end{equation}
where ${\bbox{\phi}}$ has been rescaled by a factor of $\sqrt{g}$.
Here we shall want to allow for the velocities for the different
modes to be different in which case we have to generalize the above
equation for ${\bf l}$ to obtain
\begin{equation}
{\bf l}=(\frac{1}{v_z}\phi_y\dot{\phi_z}-\frac{1}{v_y}\dot{\phi_y}\phi_z,
\frac{1}{v_x}\phi_z\dot{\phi_x}-\frac{1}{v_z}\dot{\phi_z}\phi_x,
\frac{1}{v_y}\phi_x\dot{\phi_y}-\frac{1}{v_x}\dot{\phi_x}\phi_y).
\label{eq:l}
\end{equation}
A simple mean field theory in the
spirit of the Landau-Ginzburg model can then be obtained
by considering the free model with $\alpha=0$.
This we shall refer to as the free boson model.

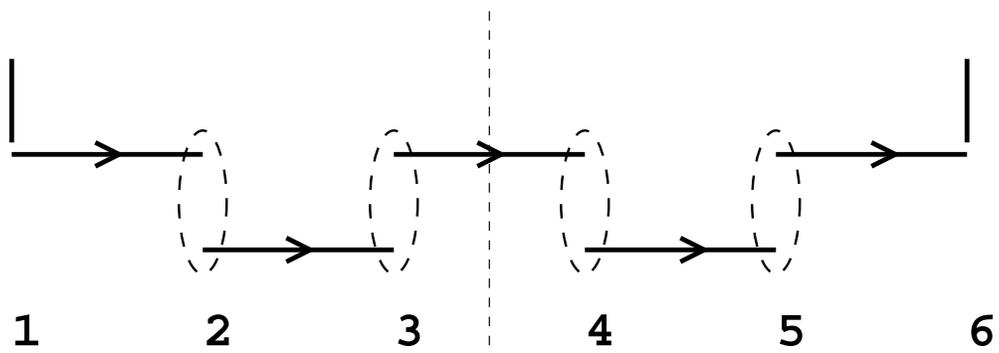
\begin{figure}
\caption{Schematic drawing of the valence bonds showing
the two $S=1/2$ end-excitations.}
\label{fig:vb}
\end{figure}

The DMRG method that we shall use applies best to systems
with open boundaries. The open boundaries has the effect of leaving
a $S=1/2$ degree of freedom at each end of the chain, see
Fig.~\ref{fig:vb}, which will interact with the rest of the system.
The open boundaries will thus introduce an interaction
of the following form
\begin{equation}
H_I=-\lambda[s\sqrt{g}{\bbox{\phi}}(1)\cdot{\bf S}^{\prime}_1+
a{\bf l}(1)\cdot{\bf S}^{\prime}_1-
s\sqrt{g}{\bbox{\phi}}(L)\cdot{\bf S}^{\prime}_L+
a{\bf l}(L)\cdot{\bf S}^{\prime}_L ],
\label{eq:il}
\end{equation}
where ${\bf S}^\prime_1$ and ${\bf S}^\prime_L$ are two $S=1/2$ excitations
known to exist at the end of the open chain~\cite{endex}.
Here we have implicitly assumed that the coupling to the staggered and
uniform magnetization can be described by one coupling
$\lambda$, we shall, however, take $\lambda$ to be different for the
in-plane and out of plane couplings, i.e. $\lambda_{\perp}\neq\lambda_{||}$
and equivalently $g_\perp\neq g_{||}$. In the following we shall
therefore explicitly write $g_a, \lambda_a$ where necessary.
The individual spins, ${\bf S}_i$, can now
be represented as
\begin{equation}
{\bf S}_i\approx {s}\sqrt{g}(-1)^{i-1}{\bbox{\phi}}+a{\bf l}
+\delta_{i,1}{\bf S}^\prime_1+\delta_{i,L}{\bf S}^\prime_L,
\label{eq:si}
\end{equation}
with ${\bf l}$ as in Eq.~(\ref{eq:l}).
In the following we sometimes set the lattice spacing, $a=1$.
An alternative field theory treatment 
of the $S=1$ Heisenberg chain using Majorana
fermions has also been proposed~\cite{tsvelik}. This model
predicts rather complicated forms for the correlation functions
and we shall therefore use the conceptually simpler free boson model.

\section{Energy Gaps}\label{sec:gaps}
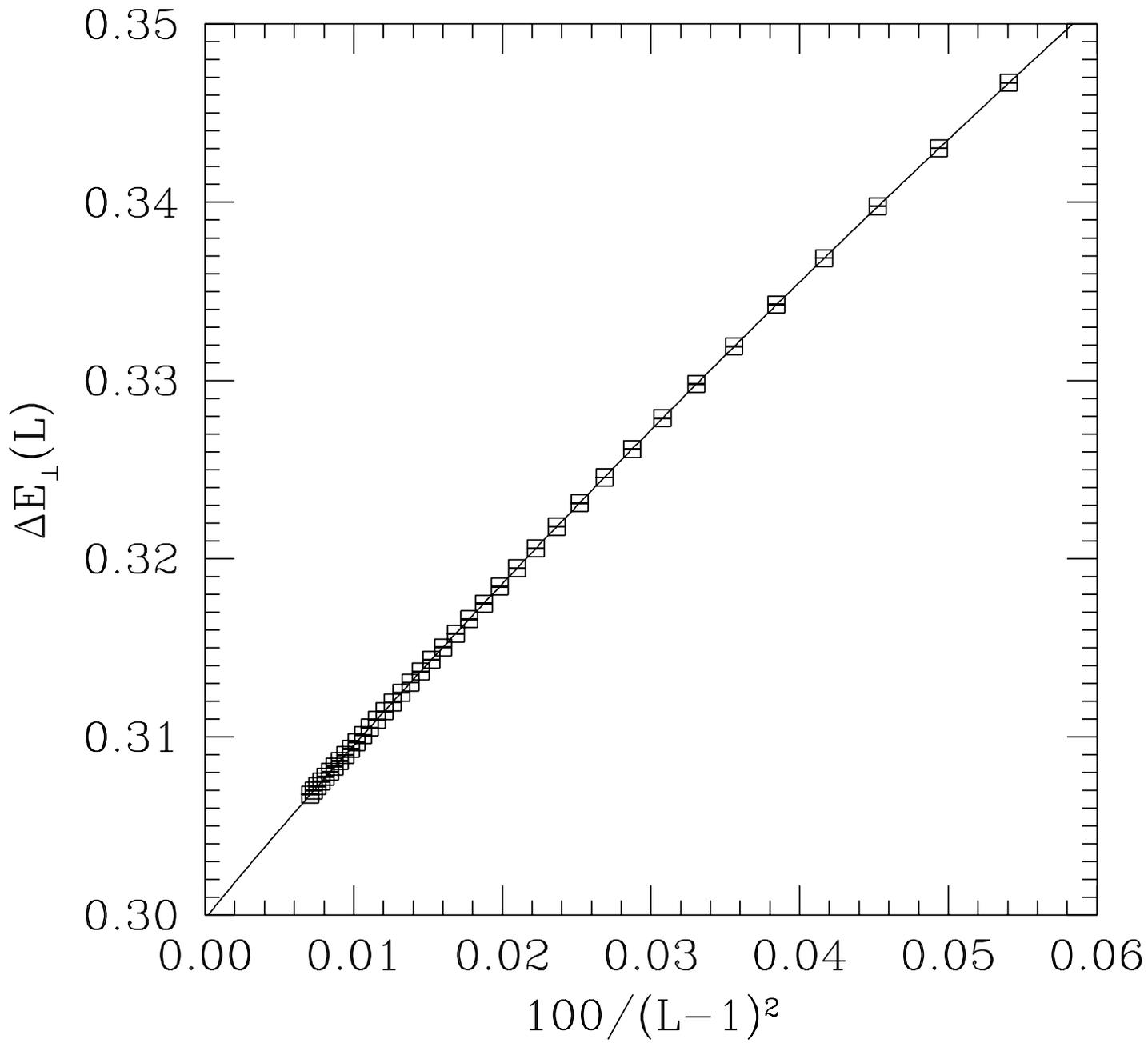
\begin{figure}
\caption{The gap $\Delta_\perp$ as a function of $100/(L-1)^2$.
The solid line indicates a best fit to the theoretically expected
form.
}
\label{fig:lgap}
\end{figure}

Due to the presence of the $D$ term in Eq.~(\ref{eq:h}) the Haldane gap
will be split into a low-lying doublet, $\Delta_{\perp}$, and a
higher-lying
singlet, $\Delta_{||}$. These two magnon modes have quantum numbers of
\begin{equation}
|a>\equiv\sum_x e^{ikx}S_x^a|0>,
\end{equation}
where $a=z$ for the singlet, and $a=x$ or $y$ for the
doublet and $|0>$ is the singlet ground-state. Thus the singlet,
$|z>$, has total $S^z$ component, $S_T^z=0$, whereas the doublet,
$|\pm>=(|x>\pm i|y>)/\sqrt{2}$, has $S_T^z=\pm 1$.
The chains that we consider here are subject to open boundary conditions
and we shall choose the exponentially low-lying $1^-$ state as our
reference state. In order to determine the two gaps we shall use
a procedure similar to what was recently used for the isotropic
chain~\cite{us1}. The addition of a bulk magnon changes the
parity~\cite{us1}. 
This can be seen as follows.
We only consider
chains of even length.
For these chains
the ground-state is a singlet with even parity, $0^+$. Above the
ground-state is an exponentially low-lying triplet, $1^-$. In the
thermodynamic
limit the triplet and the singlet become degenerate and the
ground-state four-fold degenerate. This spectrum
can be seen to arise from the two
$S=1/2$ end-excitations
forming either an odd parity singlet or an even
parity triplet, in addition to an overall parity flip coming from the
rest of the ground-state. This
parity-flip can be understood
from the valence bond solid state~\cite{jphysc}
where we draw two valence bonds emanating
from each site. These valence
bonds represent singlet contractions of pairs of
$S=1/2$'s so they have a directionality associated with them. When we
make a parity transformation we flip the orientation of an odd number
of valence bonds resulting in a $(-)$ sign.
This is schematically depicted in Fig.~\ref{fig:vb}.
Thus, the parity, $P_E$, of a
state with no magnons present
is $(+)$ if the end-excitations combine into
the singlet and $(-)$ for the triplet.
The parity of higher excited states,
containing one or more magnons, is a product of three factors,
$P_EP_{SW}P_m$.
$P_m$ contains a contribution of $(-)$ from each magnon
present.
This is because the magnons are created and annihilated by the
staggered magnetization operator, and this changes sign upon switching
even and odd sublattices. $P_{SW}$ is the parity of the spatial
wave-functions of the magnons.
The gap to the doublet should therefore be calculated
with respect to a state that has $S_T^z=\pm1$ and parity $(-1)$ as
compared to the reference state $1^-$. One candidate is
thus the state $2^+$. As shown in Ref.~\onlinecite{us1} we expect the
$L$ dependence of this gap to be 
\begin{equation}
\Delta_\perp(L)=\Delta_\perp+\frac{(v_\perp\pi)^2}{2\Delta_\perp(L-1)^2}+O(L^{-3}).
\label{eq:el}
\end{equation}
Our results are shown in Fig.~\ref{fig:lgap} where the solid line
indicates the best fit to the above form
$0.2998(1)+105.2(1)(L-1)^{-2}-794(2)(L-1)^{-3}$. From this we can
estimate $\Delta_\perp$ and $v_\perp$
\begin{equation}
\Delta_\perp=0.2998(1),\ \ v_\perp=2.53(1).
\end{equation}
The value of this gap previously obtained by Golinelli
al.~\cite{golinelli3} of $\Delta_\perp=0.301$ is in good agreement
with the above result. $v_\perp$ can be compared to the value
of $v_\perp\simeq 2.53$ that can be extracted from exact
diagonalization results~\cite{tak93} of the energy as a function
of $k$ with $D=0.2$. Presumably the dependence
of $v$ on $D/J$ is fairly small.
	
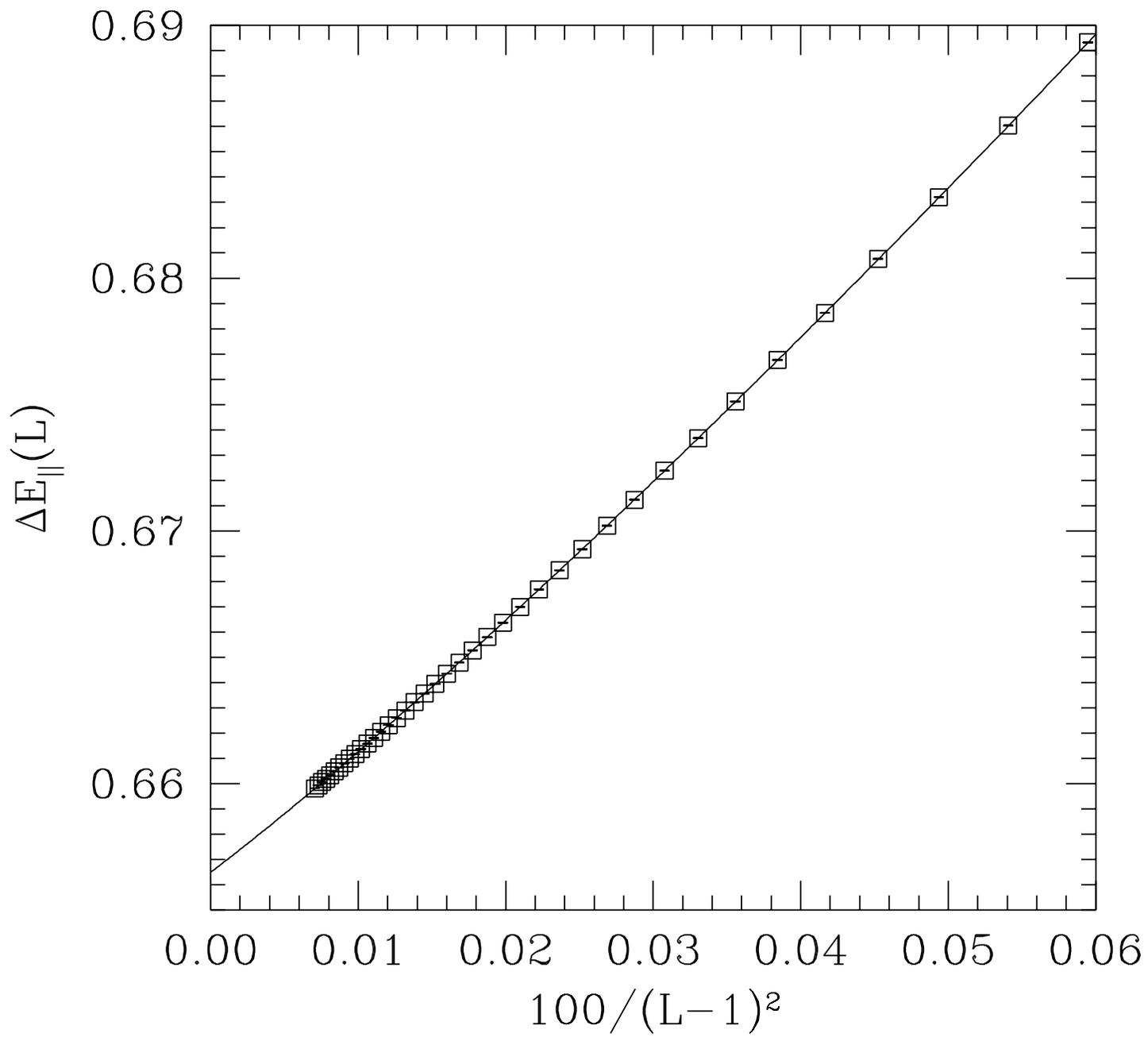
\begin{figure}
\caption{The gap $\Delta_{||}$ as a function of $100/(L-1)^2$.
The solid line indicates a best fit to the theoretically expected
form.
}
\label{fig:hgap}
\end{figure}

The gap to the heavy magnon, $\Delta_{||}$, is more difficult to
obtain. Following the above reasoning we use one of the excited
states in the $1^+$ sector to calculate $\Delta_{||}$. The expected 
dependence on $L$ is
\begin{equation}
\Delta_{||}(L)=\Delta_{||}+\frac{(v_{||}\pi)^2}{2\Delta_{||}(L-1)^2}+O(L^{-3}).
\end{equation}
We show our results in Fig.~\ref{fig:hgap} where the solid
line indicates the best fit to the above form
$0.6565(5)+42.5(1)(L_1)^{-2}+521(2)(L-1)^{-3}$. We can now
estimate $\Delta_{||}$ and $v_{||}$
\begin{equation}
\Delta_{||}=0.6565(5),\ \ v_{||}=2.38(1).
\end{equation}
The value of $\Delta_{||}$ is in good agreement with previous 
results~\cite{golinelli3}. Extracting $v_{||}$ from 
exact diagonalization results~\cite{tak93}
with $D/J=0.2$ we obtain $v_{||}\simeq2.40$ which
supports our finding of $v_{||}=2.38(1)$ for $D/J=0.18$.

For periodic systems the energies will approach their
asymptotic value exponentially fast. It is then easy
to boost results for small chain lengths, to obtain
rather good estimates of the asymptotic value, by
eliminating the exponential corrections through
the application of Shanks transformations~\cite{barber}.
For the open ended chains, that we consider here, the corrections
are powers in $L$ to leading order given by Eq.~(\ref{eq:el}).
In this case it is also possible to apply somewhat more
complicated algorithms to extract asymptotic values for the
gaps. We applied the alternating $\epsilon$-algorithm~\cite{barber}
to our data and obtained values for the gaps in complete agreement
with the above quoted values obtained by fitting to
Eq.~(\ref{eq:el}). In general these techniques seems to yield
a somewhat smaller accuracy for open boundary 
conditions~\cite{barber,betsuyaku,norbert} as compared
to periodic systems.

Using the DMRG method it is quite easy to extract the ground-state
energy per spin in the thermodynamic limit by
considering the quantity~\cite{white1,white2} $(E_{L}-E_{L-4})/4$.
We obtain $e_0/J=1.2856861$.

\section{Bulk Correlations}\label{sec:bulkcor}
We now turn to a discussion of
the bulk correlations,
$<S_i^a S_j^a>$, where both i and j are far away
from the boundaries. For the $S=1$ systems we find using
Eq.~(\ref{eq:si})
\begin{equation}
<S_i^aS_j^a>=(-1)^{i+j}g_a<\phi^a(x_i)\phi^a(x_j)>+
a^2<l^a(x_i)l^a(x_j)>.
\label{eq:bulkcorr}
\end{equation}
Here $<\cdot>$ denotes ground-state expectation values.
Let us start with the first term. 
We shall follow the approach of Ref.~\onlinecite{aw}.
We expand the staggered magnetization field,
$\bbox{\phi}$, using a mode-expansion in the magnon operators
${\bf a}_k, {\bf a}_k^{\dag}$. We use the relativistic normalization
of these operators
\begin{equation}
[a_k^a,a_{k'}^{a,\dag}]=4\pi v_a \omega_k \delta(k-k'),
\label{eq:comrel}
\end{equation}
where $\omega_k =\sqrt{\Delta^2_a+(v_ak)^2}$. 
With the definitions ${\bf X}=({\bf X_0, X_1})\equiv(t,x/v)$,
${\bf K}=({\bf K_0, K_1})\equiv(\omega,vk)$ and
${\bf K\cdot X}= \omega t-kx$, we write for
the mode-expansion
\begin{equation}
\bbox{\phi}(x,t)=\int \frac{dk}{4\pi\omega_k}
\left(e^{-i{\bf K\cdot X}}{\bf a}_k+e^{i{\bf K\cdot X}}{\bf a}_k^{\dag}\right).
\label{eq:modexp}
\end{equation}
Using the mode expansion together with the commutation relations
Eq.~(\ref{eq:comrel}) we obtain
\begin{equation}
g_a<0|\phi^a(x,0)\phi^a(0,0)|0>=g_av_a\int\frac{dk}{4\pi}
e^{-ikx}\frac{1}{\sqrt{\Delta_a^2+v_a^2k^2}}.
\label{eq:stbes}
\end{equation}
The integral can be expressed in a simple form using the
modified Bessel function $K_0$, and we obtain
\begin{equation}
g_a<0|\phi^a(x,0)\phi^a(0,0)|0>=
\frac{g_a}{2\pi}K_0\left(x/\xi_a\right)
\stackrel{\longrightarrow}{\scriptscriptstyle |x|\rightarrow\infty}\frac{g_ae^{-|x|/\xi_a}}{2\sqrt{2\pi|x|/\xi_a}},
\label{eq:basymp}
\end{equation}
where we have set $\xi_a=v_a/\Delta_a$.

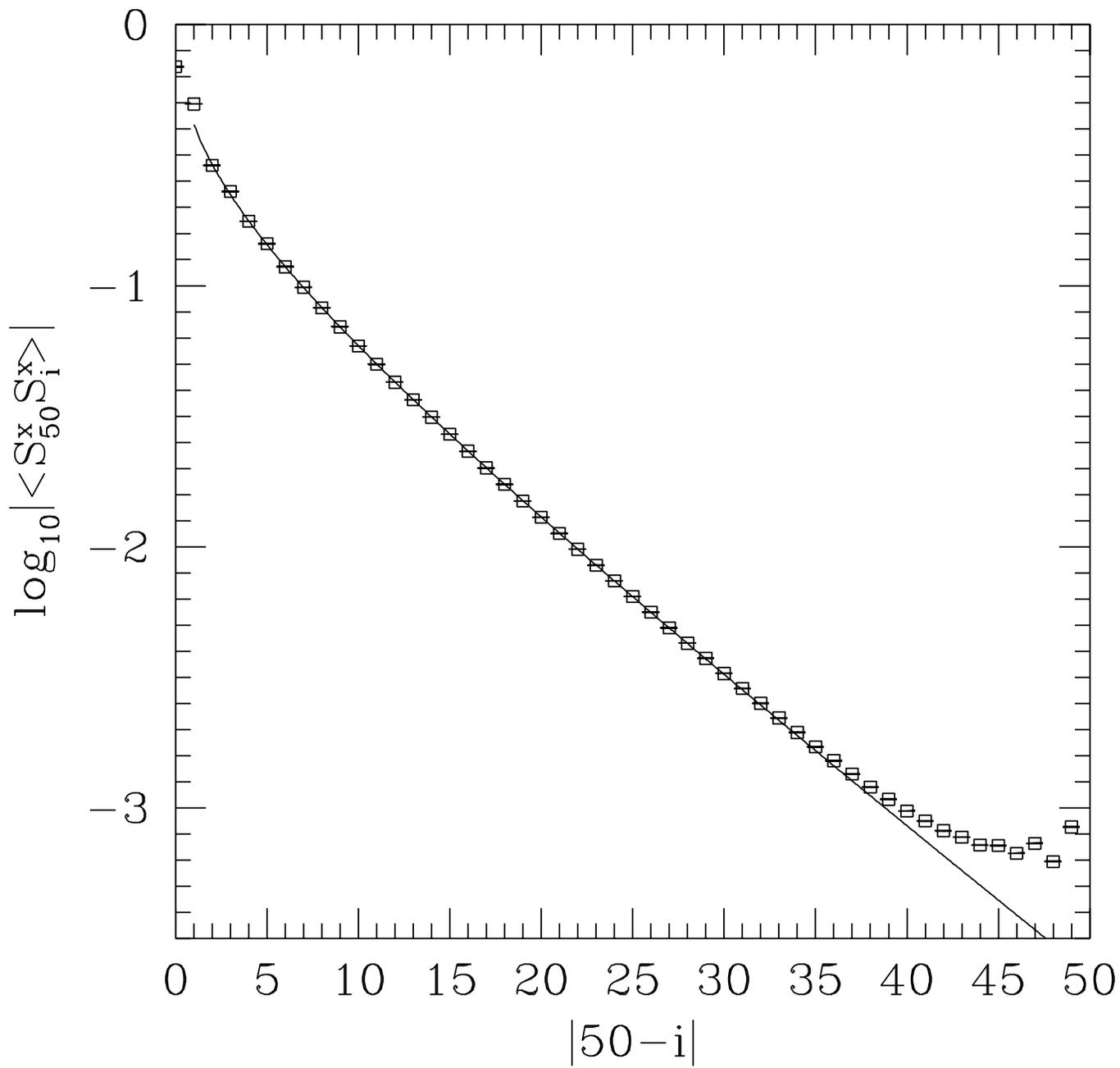
\begin{figure}
\caption{The bulk correlation function, $<S^x_iS^x_j>$, for a 100 site
{\it anisotropic} chain. Shown is $<S^x_{50}S^x_i>$ as a function
of $|50-i|$. The center of the chain is thus at the left hand
side of the figure while the chain end is approached to the right.
The solid line represents a fit to the Bessel function
form described in the text.
}
\label{fig:bulkx}
\end{figure}

Due to the factor of 2 in the size of the correlation
lengths only the leading Bessel function behavior,
given by $K_0$, can be observed for the perpendicular bulk
correlation functions, $<S^x_iS^x_j>$ and $<S^y_iS^y_j>$, as we shall see below.
For $<S^x_iS^x_j>$ we find by fitting to the expected form
\begin{equation}
<S^x_iS^x_j>=(-1)^{x}0.1843(2)K_0\left(x/8.345(8)\right),
\label{eq:bxfit}
\end{equation}
with $\chi^2=2.81$.
The results
are shown in Fig.~\ref{fig:bulkx} where the solid line
is Eq.~(\ref{eq:bxfit}). This fit determines the 
correlation length $\xi_\perp$,
\begin{equation}
\xi_\perp=8.345(8).
\end{equation}
The numerical values for this correlation function is listed in the
appendix.
In Fig.~\ref{fig:bulkx} it is also clear that the
above form fails when the chain end is approached
at the right hand side of the figure.
We see that the relation $\xi_\perp=v_\perp/\Delta_\perp$
is obeyed to within $1\%$ if one uses the values for $v_\perp$ and
$\Delta_\perp$
obtained in section~\ref{sec:gaps}. An estimate of $g_\perp$ can
also be extracted, $g_\perp\simeq1.16$. Note that the Lorentz
invariant form of Eq. (\ref{eq:bxfit}) works even for $x\leq\xi_\perp$
where the asymptotic form of Eq.~(\ref{eq:basymp}) is not valid.

The remaining term, $a^2<l^a(x_i)l^a(x_j)>$
constitutes minute corrections to the overall form given by
the Bessel function $K_0$. 
$l^z(x,t)$ contains four terms with two-magnon creation
and annihilation operators. 
The only contribution to $<l^z(x_i)l^z(x_j)>$ in the ground-state
will come from the
double creation term, $l^{z}_c$.
Thus the term $a^2<l^a(x_i)l^a(x_j)>$ in the
correlation function effectively corresponds to two magnon
excitations. From the mode expansion we see that
\begin{equation}
l^{z}_c(x,t)=i\int\frac{dk'dk''}{16\pi^2v_{\perp}\omega_{k'}\omega_{k''}}
(\omega_{k'}-\omega_{k''})e^{i({\bf K'}+{\bf K''})\cdot {\bf X}}
a_{k'}^{x\dag}a_{k''}^{y\dag}.
\label{eq:lc}
\end{equation}
In this equation $v_\perp=v_x=v_y$.
We then see, through the use of the commutation relations, that
\begin{equation}
a^2<0|l^z(x,0)l^{z}(0,0)|0>
=a^2\int\frac{dk'dk''}{16\pi^2\omega_{k'}\omega_{k''}}
(\omega_{k'}-\omega_{k''})^2e^{i(k'+k'')x}.
\end{equation}
Rearranging the integral we obtain
\begin{equation}
=\frac{a^2}{2}\left[\int\frac{dk'}{2\pi}\omega_{k'}e^{ik'x}
\int\frac{dk''}{2\pi}\frac{1}{\omega_{k''}}e^{ik''x}-\delta(x)^2\right].
\end{equation}
With our definitions the last of the two integrals is
equal to $K_0(x\Delta_\perp/v_\perp)/(\pi v_\perp)$.
The first of the two integrals
can be evaluated through the relations
\begin{equation}
\int\frac{dk'}{2\pi}\omega_{k'}e^{ik'x}=
\frac{\Delta^2_{\perp}}{v_{\perp}2\pi}\int du \sqrt{1+u^2}e^{iux/\xi_\perp}=
\frac{\Delta^2_{\perp}}{v_{\perp}\pi}(-\frac{d^2}{dx^2}+1)K_0(x/\xi_\perp),
\end{equation}
where we have used the relation $\xi_\perp=v_\perp/\Delta_\perp$.
By applying the well known recurrence relations for the
modified Bessel functions it is easy to show that
\begin{equation}
(-\frac{d^2}{dx^2}+1)K_0(x/\xi_\perp)=-\frac{1}{|x/\xi_\perp|}K_1(x/\xi_\perp),
\end{equation}
where $K_1(x)= - dK_0/dx$.
Putting our results together
we find that
\begin{equation}
a^2<0|l^z(x,0)l^{z}(0,0)|0>=-\frac{a^2}{2\pi^2\xi_\perp^2}
K_0(x/\xi_\perp)\frac{1}{|x/\xi_\perp|}K_1(x/\xi_\perp), \ \ x\neq 0.
\label{eq:ll}
\end{equation}
One should note that $K_1$ is positive definite and the 
contribution is thus negative. 
This term will asymptotically behave as $\sim -(1/x^2)\exp(-2x/\xi_\perp)$.
Since $2/\xi_\perp\sim\xi_{||}$ the exponential dependence is of the same
order as the leading term given by $K_0$. We have evaluated this term
in the free boson model neglecting interaction effects, these could
be of importance but are unlikely to change the asymptotic behavior
of the two magnon term (see also section~\ref{sec:struct}). For the
$<S^x_iS^x_j>$ correlation function this last term has a
more complicated form with an exponential decay faster than the
leading term and it is therefore more difficult to observe.

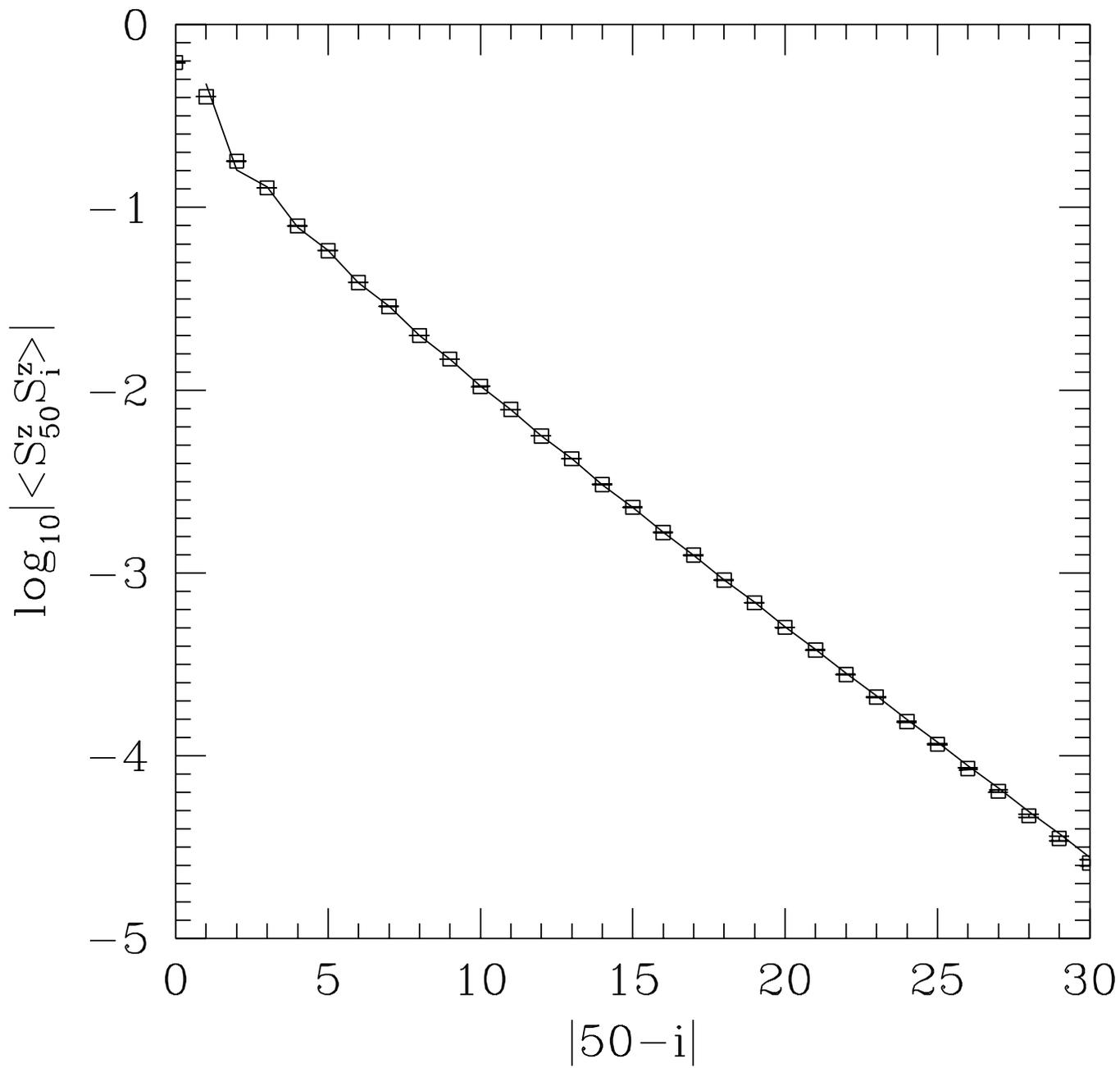
\begin{figure}
\caption{The bulk correlation function, $<S^z_iS^z_j>$, for a 100 site
{\it anisotropic} chain. Shown is $<S^z_{50}S^z_i>$ as a function
of $|50-i|$. The center of the chain is thus at the left hand
side of the figure while the chain end is approached to the right.
The solid line connects the discrete points obtained by
fitting to the form described in the text.
}
\label{fig:bulkz}
\end{figure}

Using these results we now consider the $<S^z_iS^z_j>$
correlations. From the above we have
\begin{equation}
<S^z_iS^z_j>\simeq(-1)^{x}\frac{A}{2\pi}K_0\left(x/\xi_{||}\right)
-\frac{B}{2\pi^2\xi_\perp^2}K_0\left(x/\xi_\perp\right)\frac{2}{|x/\xi_\perp|}
K_1\left(x/\xi_\perp\right).
\label{eq:zcor}
\end{equation}
Here $x=|i-j|$, $A=g_{||}$, and $B=1$ from the above derivation.
Since $\xi_\perp$ is roughly twice $\xi_{||}$ in NENP there is a
chance the last term is detectable in the
$<S^z_iS^z_j>$ correlation
function.
In Fig.~\ref{fig:bulkz} is shown $<S^z_{50}S^z_i>$ as a function
of $|50-i|$. Fitting to the above form we find
\begin{eqnarray}
<S^z_iS^z_j>=& &(-1)^{x}0.2178(6)K_0\left(x/3.69(1)\right)\nonumber\\
& &-0.007(3)K_0\left(x/8.9(9)\right)\frac{1}{|x|}K_1\left(x/8.9(9)\right),
\label{eq:bzfit}
\end{eqnarray}
with $\chi^2=6.25$.
The solid line in Fig.~\ref{fig:bulkz} connects the discrete
points obtained from Eq.~(\ref{eq:bzfit}).
The numerical values for this correlation function is listed in the
appendix.
This fit determines $\xi_{||}$, 
\begin{equation}
\xi_{||}=3.69(1).
\end{equation}
The relation $\xi_{||}=v_{||}/\Delta_{||}$ is then satisfied to within
$2\%$ if one uses the values for $v_{||}$ and $\Delta_{||}$
obtained in section~\ref{sec:gaps}. In addition we also find
$g_{||}\simeq 1.37$. We can also compare the fitted coefficient
for the two-magnon term, 0.007, with the theoretical prediction
of $a^2/(2\pi^2\xi_\perp)$. Setting $a=1$ and using the
previously obtained estimate for $\xi_\perp$ we see that the
prediction is $\sim0.006$ in good agreement with the
fitted value. The value we find for $\xi_\perp$ is also
in reasonable agreement with what we previously obtained by
looking directly at the $<S^x_iS^x_j>$ correlation function.

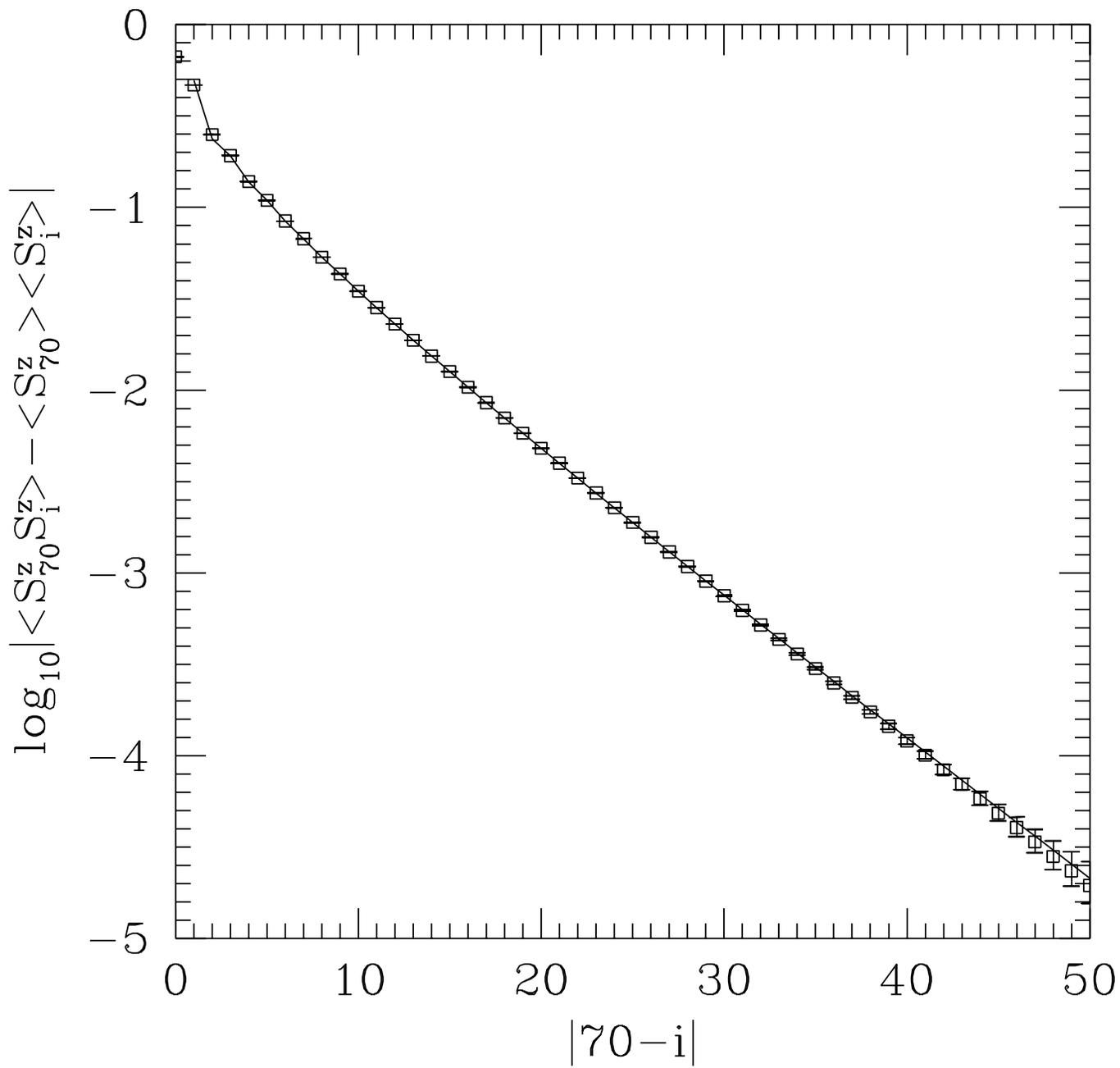
\begin{figure}
\caption{The bulk correlation function, $<S^z_iS^z_j>$, for a 100 site
{\it isotropic} chain.
Shown is $<S^z_{70}S^z_i>-<S_{70}^z><S_i^z>$ as a function
of $|70-i|$. 
The chain end is approached to the right.
The solid line connects the discrete points obtained by
fitting to the form described in the text.
}
\label{fig:biso}
\end{figure}

In the case of the isotropic chain, where $D=0.0$, we also expect 
the bulk correlation functions to behave as Eq.~(\ref{eq:zcor}). Our 
results are shown in Fig.~\ref{fig:biso} for $<S^z_{70}S^z_i>$. Since
the calculation is performed in the $1^-$ state we expect
a disconnected term of the form $<S^z_{70}><S^z_i>$ which we subtract
from $<S^z_{70}S^z_i>$. Such a disconnected term is also present when
$D=0.18$ but is orders of magnitude smaller due to the smaller
correlation length $\xi_{||}$ and due to the fact that $S^z_{50}$ is
right at the middle of the chain whereas $S^z_{70}$ is much
closer to the boundary for a 100 site chain.
For small arguments in $<S^z_{j}S^z_i>$ we expect the same effect
to be present in the $0^+$ state since the wave function locally looks
the same; only by considering large arguments will one be able 
to differ between the two states.
Fitting $<S^z_{70}S^z_i>-<S^z_{70}><S^z_i>$ to Eq.~(\ref{eq:zcor})
we obtain
\begin{eqnarray}
<S^z_iS^z_j>-<S^z_{i}><S^z_j>
=& &(-1)^{x}0.1999(1)K_0\left(x/6.03(1)\right)\nonumber\\
& &-0.008(5)K_0\left(x/7(2)\right)\frac{1}{|x|}K_1\left(x/7(2)\right),
\end{eqnarray}
with a $\chi^2=20$. The fit works very well even for rather small values
of $|i-70|$.
From this fit we can determine $\xi=6.03(1)$ (in complete agreement
with Ref. \onlinecite{white2}), and $g=1.26$.
The numerical values for this correlation function is listed in the
appendix.
For the isotropic Heisenberg chain the best numerical estimates for
$\Delta,\ v,\ \xi$ are~\cite{white2,us1}
$\Delta=0.41050(2),\ v=2.49(1),\ \xi=6.03(1)$,
which shows that the relation $\xi\sim v/\Delta$ is correct to
within $0.5\%$.
Again we can compare the coefficient in front of the double
Bessel function term to what we expect from the free boson theory,
$1/(2\pi^2\xi)\sim0.008$, in very good agreement with the
fitted value. The double Bessel function term is in this case
rather difficult to fit due to the fact that it is exponentially
small compared to the leading $K_0$ term. The value we find for 
the correlation length from this term has therefore rather large
error-bars as do the coefficient in front of it. Both are, however,
in good agreement with the free boson theory.
The leading Bessel function
behavior, $K_0$, of the bulk correlation functions has previously
been determined for the isotropic
case~\cite{white2,nomura,liang,meshkov1}.

The values $g_\perp\simeq1.16$ and $g_{||}\simeq 1.37$
obtained above gives an average $\bar{g}=1.27$ which
compares well with what we obtain for the isotropic chain,
$g=1.26$. The value of $g$ for the isotropic chain
can roughly be compared to
the value of $g\simeq 1.44$ for the bare coupling
obtained~\cite{chubokov} for the isotropic
chain by $1/S$ expansion. 

\section{End Effects}\label{sec:boundcor}
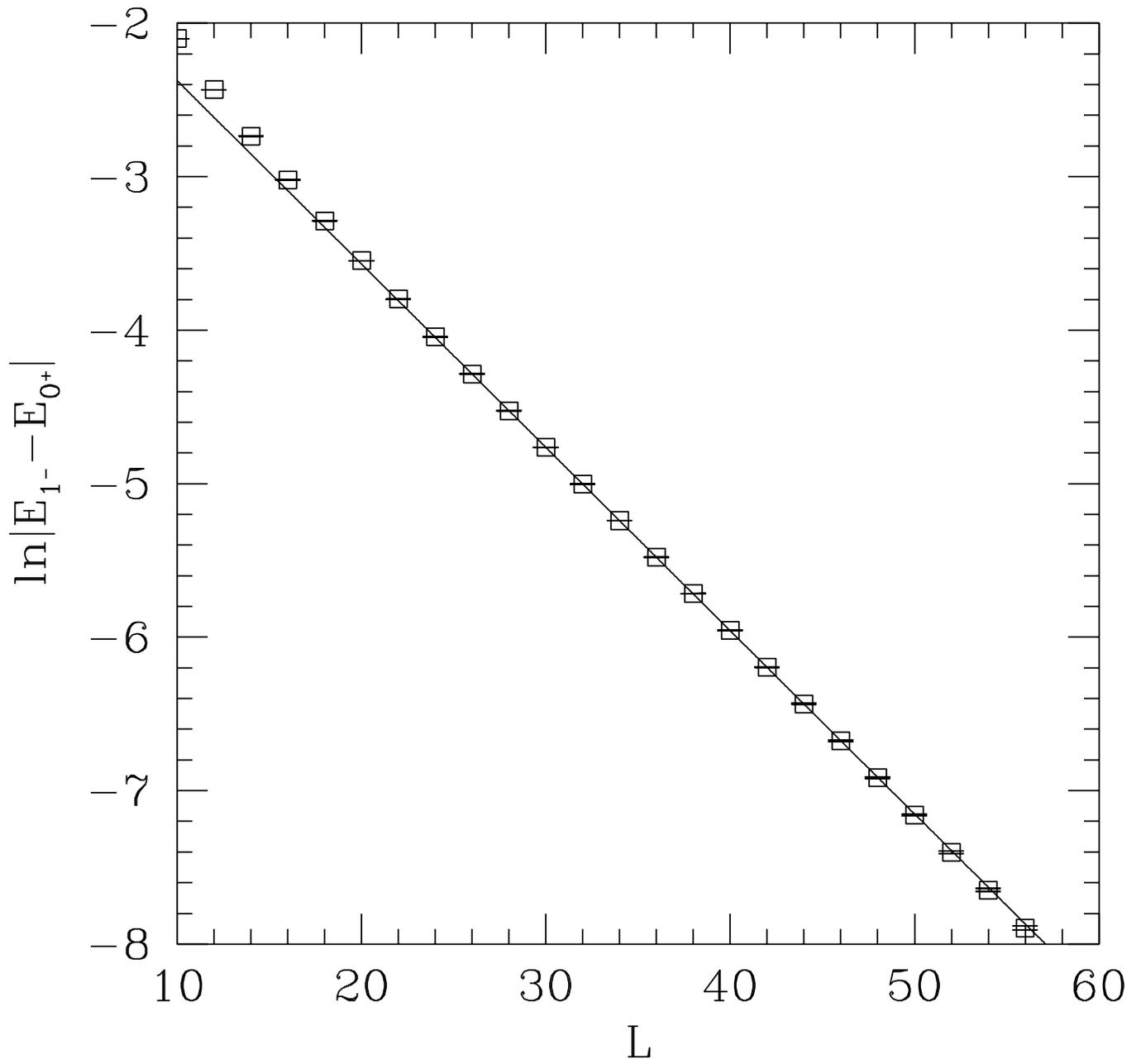
\begin{figure}
\caption{The gap, $E_{1^-}-E_{0^+}$, as a function of
chain length. The dependence is exponential defining
a characteristic length of $8.38$.
}
\label{fig:gap01}
\end{figure}

As mentioned above, all of our numerical results are for chains
subject to open boundary conditions. We therefore need to consider
chain end effects. While at first sight such effects
might seem undesirable they allow consistency checks on our results.
For instance, they allow
another way of estimating $\xi_\perp$ by looking directly
at the energy levels. As mentioned, the $1^-$ state is exponentially
low-lying. The gap between the $1^-$ state and the true ground-state,
$0^+$, decays exponentially with the chain length, $L$.
The energy difference arises because the two $S=1/2$ end-excitations
are in the triplet configuration in the $1^-$ state as opposed to a
singlet for the ground-state, $0^+$. The interaction between the
$S=1/2$ end-excitations are predominantly 
determined by the exchange of virtual
magnons of the $\Delta_\perp$ type~\cite{partha}.
The energy difference should therefore
decay exponentially with a characteristic length equal
to $\xi_\perp$. This can be seen by integrating out $\bbox{\phi}$
in the quadratic Hamiltonian of Eq.~(\ref{eq:fb}), including the interaction
term between the chain-end spins and the staggered magnetization field,
${\bf \phi}$ but ignoring their coupling to ${\bf l}$ for simplicity. 
(This  leads to effects which decay more rapidly with $L$.) We obtain
an effective action $S_{\rm eff}(\bf S^{\prime}_1, \bf S^{\prime}_L)$.
Ignoring the interactions involving ${\bf l}$ in Eq.~(\ref{eq:il}) we
can write 
\begin{equation} 
\int {\cal D}\bbox{\phi}
e^{-S(\bbox{\phi})+\int d\tau \lambda\sqrt{g} \left(\bf
S^{\prime}_1\cdot\bbox{\phi}(1,\tau)
-S^{\prime}_L\cdot\bbox{\phi}(L,\tau)\right)}=e^{-S_{\rm eff}}.
\end{equation}
We then find
\begin{eqnarray}
S_{\rm eff}= -{\lambda^2\over 2}g\int d\tau_1&d\tau_2&[{\bf
S}_1^{\prime}(\tau_1)-{\bf S}_L^{\prime}(\tau_1)]\cdot [{\bf
S}_1^{\prime}(\tau_2)-{\bf S}_L^{\prime}(\tau_2)]\nonumber\\
&\times&\frac{dk d\kappa}{(2\pi)^2}
\frac{v}{\kappa^2+v^2k^2+\Delta^2}e^{i(k(L-1)-\kappa (\tau_1-\tau_2))},
\end{eqnarray}
where $\kappa$ is the imaginary frequency. By Taylor expanding ${\bf
S}_{1,L}^{\prime}(\tau_2)$ around $\tau_2\approx \tau_1$, we obtain a
series of instantaneous terms involving increasing numbers of $\tau$
derivatives.  The lowest order term gives, after doing one $\tau$
integral,  an effective Hamiltonian:
\begin{equation}
H_{\rm eff}= \lambda^2g{\bf S}_1^{\prime}\cdot
{\bf S}_L^{\prime}
\int\frac{dk}{2\pi}
\frac{v}{v^2k^2+\Delta^2}e^{ik(L-1)}.
\end{equation}
The resulting integral now yields the familiar exponential
form and we can write,
\begin{equation}
H_{\rm eff}= \lambda^2 g{\bf S}_1^{\prime}\cdot
{\bf S}_L^{\prime}
\frac{e^{-(L-1)/\xi}}{2\Delta},
\label{eq:heff}
\end{equation}
where $\xi, \Delta$ and $g$ are different for the two
modes.  Note that this effective Hamiltonian is exponentially small,
for large $L$.  It follows that a time derivative of ${\bf
S}_L^{\prime}$ is of order this exponentially small energy scale. Hence
the higher derivative terms in the effective Hamiltonian will be
suppressed by powers of the ratio of this exponentially small energy
to $\Delta$ and can be safely ignored.  Due to  the large
difference between $\xi_\perp$ and $\xi_{||}$, we can write approximately
\begin{equation} 
H_{\rm eff} \simeq \lambda_\perp^2 g_\perp
\frac{1}{2}\left[S_1^{\prime +}S_L^{\prime -}+ S_1^{\prime -}S_L^{\prime
+}\right] \frac{e^{-(L-1)/\xi_\perp}}{2\Delta_\perp}. 
\end{equation}
The equivalent contribution from the
heavy magnon is smaller by a factor $\exp(-(L-1)(1/\xi_{||}-1/\xi_\perp))$,
and can therefore be ignored. We now want to compare
the $0^+$ state with the $1^-$ state. In the $0+$ state
the wave function for the two spin-1/2 end excitations is 
approximately $(1/\sqrt{2})(|\uparrow\downarrow>-|\downarrow\uparrow>)$,
while in the $1^-$ state it is just $|\uparrow\uparrow>$. 
We thus obtain
\begin{equation}
E_{1^-}-E_{0^+}=\frac{\lambda_\perp^2}{2}g_\perp
\frac{e^{-(L-1)/\xi_\perp}}{2\Delta_\perp}.
\end{equation}
In Fig.~\ref{fig:gap01} we show this energy gap on
a log scale as a function of the chain length $L$. Clearly there is
an exponential dependence and we estimate $\xi_\perp$ by a simple
linear fit to the data obtaining $\xi_\perp\sim8.38(4)$ in nice agreement
with the above result. The full functional form is given by
$E_{1^-}-E_{0^+}\simeq 0.27\exp(-(L-1)/8.38)$. Using the above 
obtained values for $\xi_\perp, \Delta_\perp$ and $g_\perp$ we find
$\lambda_\perp\simeq 0.53$.

An equivalent analysis can be done for the isotropic chain. In that
case we obtain
\begin{equation}
E_{1^-}-E_{0^+}=\frac{3\lambda^2}{4}g
\frac{e^{-(L-1)/\xi}}{2\Delta}.
\end{equation}
Fitting to the numerically determined energies we find
$E_{1^-}-E_{0^+}\simeq 0.592\exp(-(L-1)/6.07)$. The
observed correlation length is again very close to what we
found above using the bulk correlation function. The prefactor
allows us to determine $\lambda$. Using the values $g\simeq 1.26$ and
$\Delta=0.4107$, determined above, we find $\lambda\simeq0.72$.

We now consider $<{\bf S}_i>$ in the $1^-$ state.  $<{\bf S}_i^a>$ must
vanish for $a=x$ or $y$ by rotational symmetry, but can be non-zero for
$a=z$, having the value:
\begin{equation}
<S_i^z>=\delta_{i1}<S_1^{\prime z} >+\delta_{iL}<S_L^{\prime z}>
+(-1)^{i-1}\sqrt{g}<\phi^z(x_i)>+a<l^z(x_i)>.
\end{equation}
We expect $S_1^{\prime z}\approx S_L^{\prime z}\approx 1/2$ in the $1^-$
state, from $H_{\rm eff}$ of Eq.~(\ref{eq:heff}). Due to the very small energy
scale in $H_{\rm eff}$ the effective chain-end spins are ``frozen'' on
time scales of $O(\Delta^{-1})$ so we can treat them classically in
considering the response of the bulk fields, $\bbox{\phi}$ and $\bf l$. 
They act as classical sources at the two ends of the chain.  If we
temporarily ignore the couplings of the chain-end excitations to the
uniform magnetization density, $\bf l$, then we can calculate the
response of $\bbox{\phi}$ exactly, from the free boson model of
Eq.~(\ref{eq:fb})
and (\ref{eq:il}).
The field equation for $\phi^z(x_i)$  is
\begin{equation}
\left(v_{||}\frac{\partial^2}{\partial x^2}
-\frac{\Delta_{||}^2}{v_{||}}\right)\phi^z
=-\frac{\lambda_{||}\sqrt{g_{||}}}{2}[\delta (x-1)-\delta (x-L)].
\end{equation}
This equation has the solution
\begin{equation}
\phi^z_{\rm cl}(i)= \frac{\lambda_{||}\sqrt{g_{||}}}{2}
\frac{e^{-(i-1)\Delta_{||}/v_{||}}-e^{-(L-i)\Delta_{||}/
v_{||}}}{(2\Delta_{||})}, 
\end{equation}
and we see
that $|<S^z_i>|$ should decay exponentially away from the
chain ends with a correlation length $\xi_{||}=v_{||}/\Delta_{||}$. 
Note that, unlike the bulk correlation function, there is no
$1/\sqrt{|x|}$ prefactor in this case; rather than a Bessel function,
we obtain a pure exponential.

Next we consider the uniform part, $a<l^z(x_i)>$.  It is presumably not
possible to find a closed form expression for this quantity, from the
free boson model of Eq.~(\ref{eq:fb}) and (\ref{eq:il})
even after replacing the chain-end spins by their expectation values.
We will content
ourselves with doing lowest order perturbation theory in the coupling,
$\lambda$, between $l^z$ and $S_1^{\prime z}$ to make a rough estimate
of its magnitude.  This gives: 
\begin{equation}
a<l^z(x_i)>={a^2\lambda \over 2}\int d\tau <l^z(x_i,0)l^z(0,\tau)>_0,
\end{equation}
where this expectation value is calculated in the theory with $\lambda
=0$.  This is similar to what was calculated in Sec.~\ref{sec:bulkcor}
on bulk correlations in Eq.~(\ref{eq:ll}). Using the fact that
$K_0(\sqrt{x^2+v^2\tau^2}/\xi)$ obeys the relation
$(-\partial^2/\partial x^2-\partial^2/\partial
\tau^2+1)K_0(\sqrt{x^2+v^2\tau^2}/\xi)=\delta(x)\delta(\tau)$
we use Eq.~(\ref{eq:ll}) to obtain
\begin{equation}
a<l^z(x_i)>={a^2\lambda \over 4\pi^2\xi^2_\perp}\int d\tau 
K_0(\sqrt{x^2+v_\perp^2\tau^2}/\xi_\perp)\frac{\partial^2}{\partial \tau^2}
K_0(\sqrt{x^2+v_\perp^2\tau^2}/\xi_\perp).
\end{equation}
For large $x$ the Bessel function has the behavior
$(1/\sqrt{|x|/\xi_\perp})\exp(-|x|/\xi_\perp-v^2_\perp\tau^2/(2|x|\xi_\perp))$,
which results in an asymptotic decay for $a<l^z(x_i)>$
as $-(1/x^{3/2})e^{-2|x|/\xi_\perp}$.

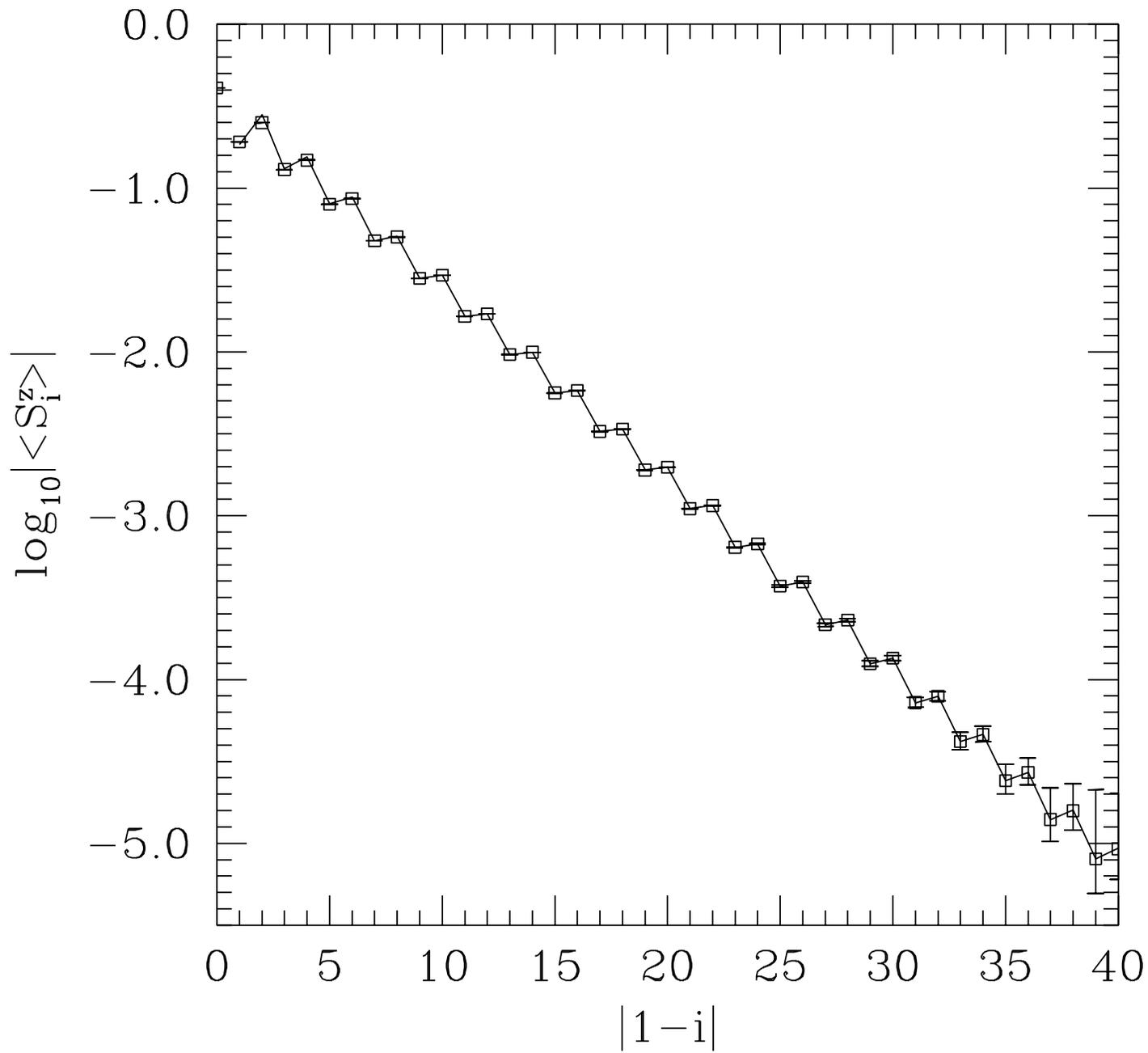
\begin{figure}
\caption{$<S^z_i>$, for a 100 site
{\it anisotropic} chain as a function
of $|1-i|$. The end of the chain is at the left hand
side of the figure.
The solid line connects the discrete points obtained from the fit
$(-1)^{i-1}0.380(2)\exp(-(i-1)/3.703(4))+0.133(4)\exp(-2(i-1)/8.45(6))
/{\protect\sqrt{i-1}}$.
}
\label{fig:sz}
\end{figure}

In Fig.~\ref{fig:sz} we show our results for $|<S^z_i>|$
as a function of $|1-i|$. The solid line connects the
discrete points obtained by a fit to the form
\begin{equation}
(-1)^{i-1}0.380(2)e^{-(i-1)/3.703(4)}+0.133(4)\frac{e^{-2(i-1)/8.45(6)}}
{\sqrt{i-1}},
\end{equation}
with $\chi^2=0.06$.
Again we see that the value for the correlation lengths,
\begin{equation}
\xi_{||}=3.703\pm0.004, \ \  \xi_{\perp}=8.45,
\end{equation}
are in excellent agreement with what we determined from the bulk
correlation functions.  The absence of any $1/x$ prefactor in the
staggered part is also in accord with the boson model.  However, the
$1/\sqrt{x}$ prefactor in the uniform part is in disagreement with the
$1/x^{3/2}$ prediction of the boson model, also the sign is wrong for that
term.  We do not understand these discrepancies. We cannot exclude the
possibility that they are somehow generated by the iterative
numerical procedure, although we find it unlikely.
From the staggered part of the fit we can
determine  $\lambda_{||}\sim 0.85$.

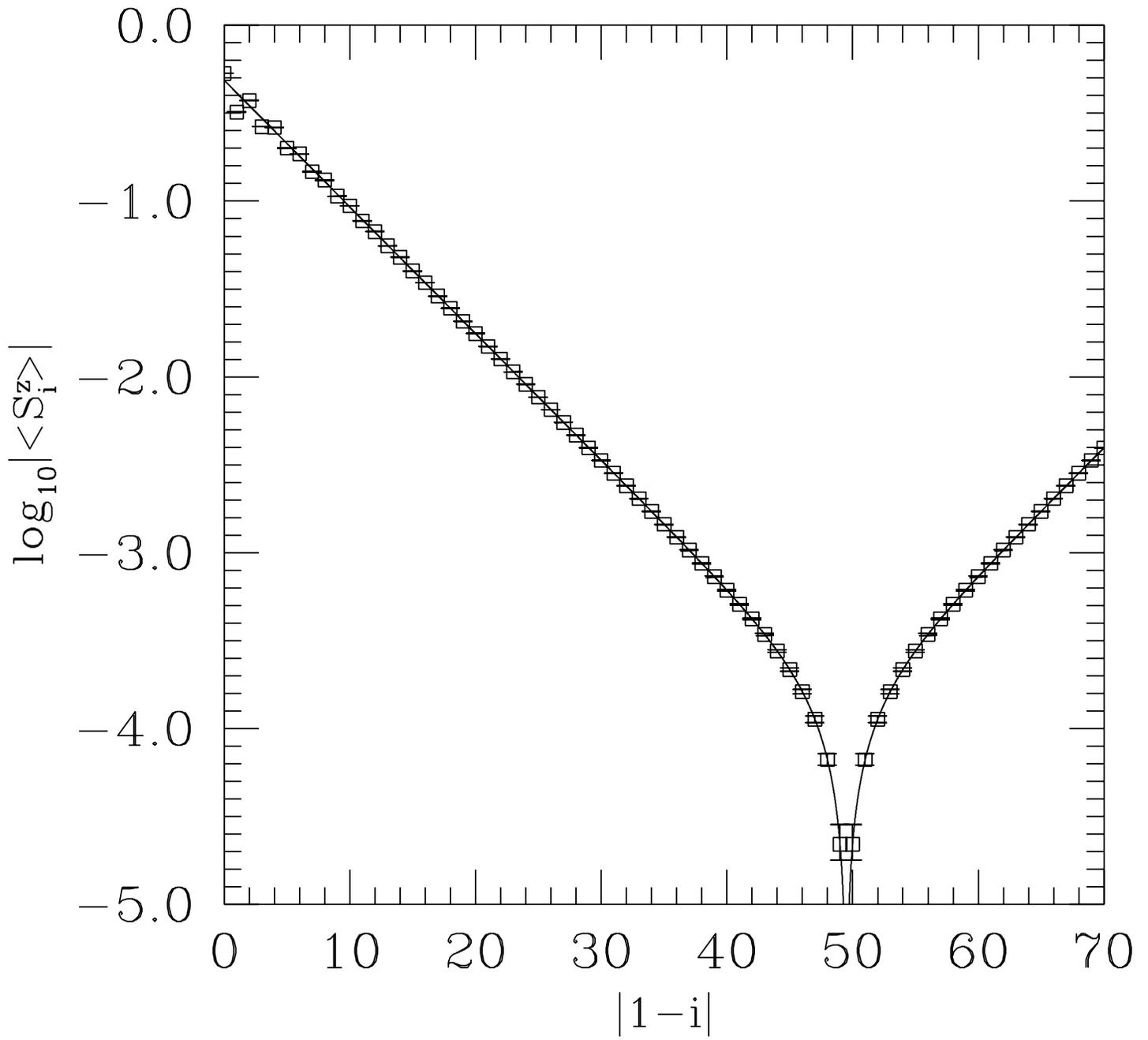
\begin{figure}
\caption{$<S^z_i>$, for a 100 site
{\it isotropic} chain as a function
of $|1-i|$. The end of the chain is at the left hand
side of the figure.
The solid line represents a fit to the points with
of the form
$0.486(1)[\exp(-(i-1)/6.028(3))-\exp(-(100-i)/6.028(3)]$.
}
\label{fig:sziso}
\end{figure}

In Fig:~\ref{fig:sziso} we show $<S^z_i>$ for a 100 site
{\it isotropic} chain. In this case we fit to an exponential form
and we obtain with a $\chi^2=6.7$,
\begin{equation}
0.486(1)[\exp(-(i-1)/6.028(3))-\exp(-(100-i)/6.028(3)].
\end{equation}
One should here
remember that the two contributions from each end of the chain add up
out of phase, hence the minus sign between the two exponential
terms. 
This clearly establishes $\xi=6.03(1)$, again in complete agreement
with our previous result and Ref.~\onlinecite{white2}.
Also we can estimate $\lambda$ to be $\lambda\sim 0.71$, which is
in excellent agreement with the value $\lambda\sim 0.72$ determined
above.

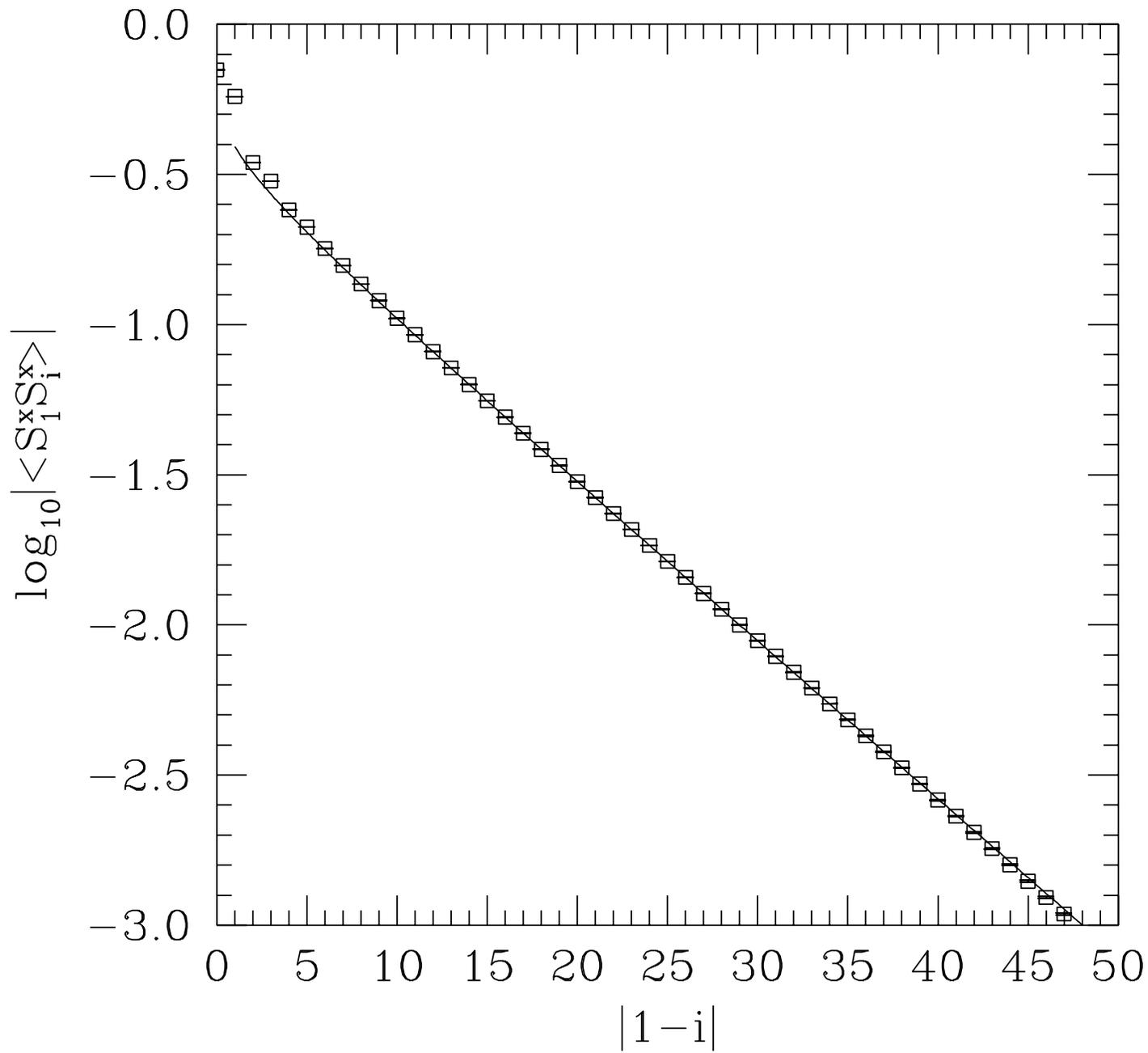
\begin{figure}
\caption{The boundary correlation function, $<S^x_1S^x_i>$, for a 100 site
{\it anisotropic} chain. Shown is $<S^x_{1}S^x_i>$ as a function
of $|1-i|$. The end of the chain is thus at the left hand
side of the figure.
The solid line represents a fit of the form
$0.277(2)\exp(-(i-1)/8.37(2))+0.064(3)K_0((i-1)/8.37(2))$.
}
\label{fig:endx}
\end{figure}

Now we consider the correlation function, $< S^a_1S^a_i>$, with one
operator at the edge of the system.  This has a large number of
contributions using the expression of Eq.~(\ref{eq:si})
for the spin operators
and the Hamiltonian of Eq.~(\ref{eq:fb}) and (\ref{eq:il}).
Let us first consider
$<S^x_1S^x_i>$, which is a bit simpler because $<S^x_i>=0$.  The
staggered part is given by:
\begin{equation}
(-1)^{i-1}[gs^2<\phi^x
(x_1)\phi^x(x_i)>+\sqrt{g}s<S^{\prime x}_1\phi^x(x_i)>].
\end{equation}
While the first term gives a similar expression to that obtained in the
bulk, the second term has a somewhat different asymptotic behavior.  
This can be extracted by doing first order perturbation theory in
$\lambda$:
\begin{equation}
<S_1^{\prime x}\phi^x(x_i)>\sim
\lambda_x{g_x}\int d\tau<\phi^x(1,\tau)\phi^x(x_i,0)>_0
<S_1^{\prime x}(\tau)S_1^{\prime x}(0)>_0.
\end{equation}
Here the Green's functions are calculated in the $\lambda =0$ limit. 
Hence, ${\bf S}^\prime_1(\tau)$ is time-independent and 
$<S_1^{\prime x}(\tau)S_1^{\prime x}(0)>_0=1/4$.  Thus:
\begin{equation}
<S_1^{\prime x}\phi^x(x_i)>\sim
\frac{\lambda_{\perp}{g_{\perp}}v_{\perp}}{4}
\int\frac{dk}{2\pi}\frac{e^{ik(x_i-1)}}{v_{\perp}^2k^2+\Delta_{\perp}^2}
=\frac{\lambda_{\perp}{g_{\perp}}}{4}
\frac{e^{-(x_i-1)\Delta_{\perp}/v_{\perp}}}{2\Delta_{\perp}}.
\end{equation}
This has a pure exponential decay, unlike the bulk part which has a
$1/\sqrt{x}$ prefactor.  The uniform part has an exponential decay of
the form $e^{-x_i(1/\xi_\perp+1/\xi_{||})}$.  This decays much more
rapidly than the staggered part and is essentially unobservable in our
numerical work.
Our data for $<S^x_1S^x_j>$ are shown in Fig.~\ref{fig:endx}.
For small arguments there is an indication of a uniform
part but we are not able to determine this. For large
arguments the points of
this boundary correlation function clearly
falls onto one line and we make a fit of the following form
\begin{equation}
<S^x_1S^x_i>\sim(-1)^{x}[0.277(2)e^{-x/8.37(2)}+0.064(3)K_0(x/8.37(2))],
\end{equation}
with $x=|i-1|$, which is shown as the solid line in Fig.~\ref{fig:endx}. 
For this fit $\chi^2=32$.
This
fit is just marginally acceptable hinting at the presence of
a uniform part.
We can again read off the correlation length
\begin{equation}
\xi_\perp=8.37(3),
\end{equation}
in excellent agreement with what we previously obtained. The coefficient
in front of the exponential should be
$\frac{\lambda_{\perp}{g_{\perp}}}{8\Delta_{\perp}}\sim 0.26$ using
our previously obtained values which is close to the fitted value.

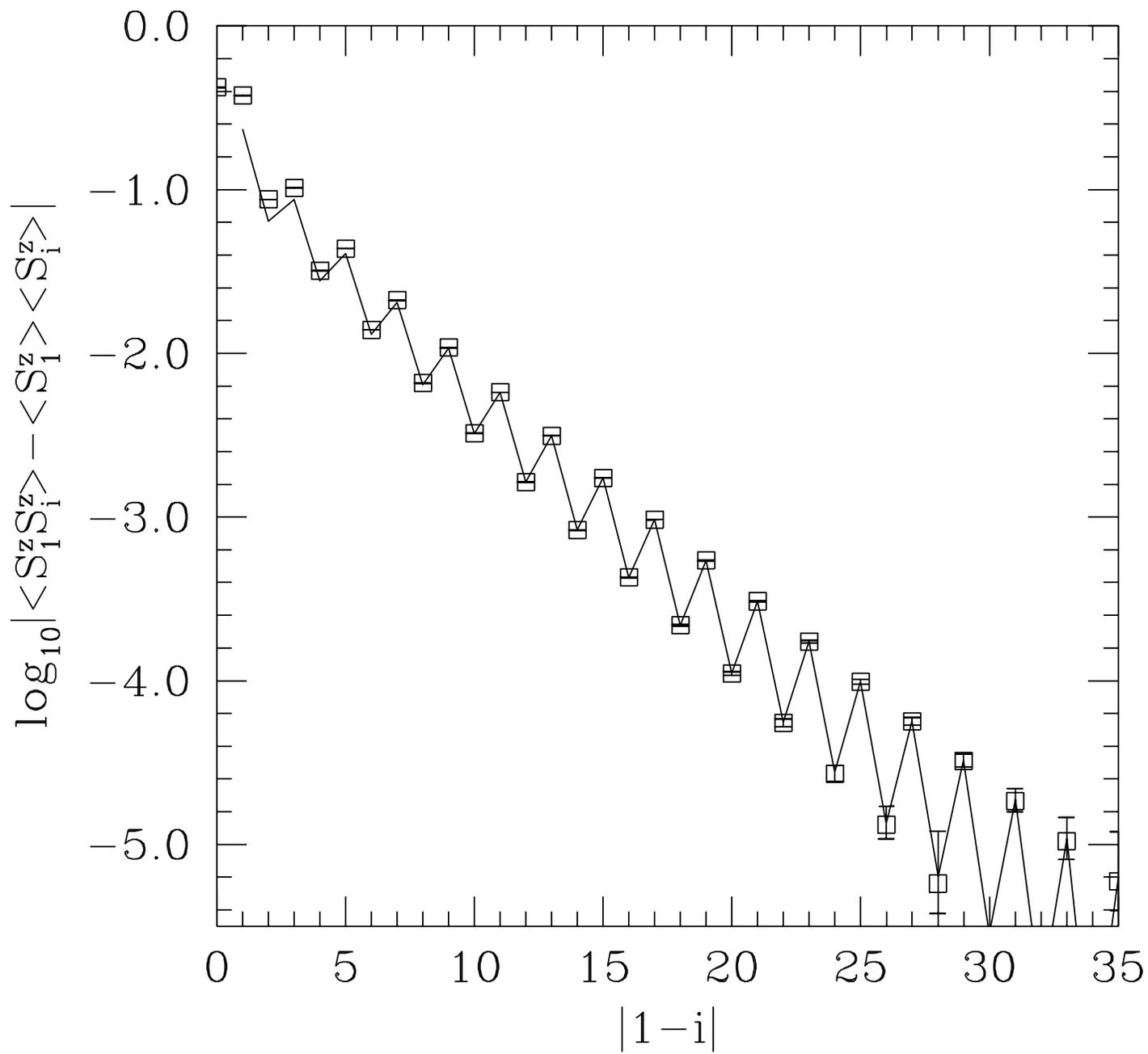
\begin{figure}
\caption{The boundary correlation function,
$<S^z_1S^z_i>-<S^z_1><S^z_i>$, for a 100 site {\it anisotropic }
chain. Shown is $<S^z_{1}S^z_i>-<S^z_1><S^z_i>$ as a function
of $|1-i|$. The end of the chain is thus at the left hand
side of the figure.
The solid line connects the discrete points obtained
by fitting to the form described in the text.
}
\label{fig:endz}
\end{figure}

The boundary correlation functions $<S^z_1S^z_i>$ is more complicated
for two reasons.  Firstly, $<S^z_i>\neq 0$ as discussed above. 
Secondly, the uniform part decays with the exponent $2x/\xi_{\perp}$
which is actually slightly {\it smaller} than the one occurring in the
staggered part, $x/\xi_{||}$.  Therefore it is also observable. We
consider the connected part: $<S^z_1S^z_i>-<S^z_1><S^z_i>$. The
staggered part is expected to have a pure exponential term as well as a
Bessel function term by similar arguments to those given above for the
boundary correlation function $<S^x_1S^x_i>$.  The uniform part is
expected to have a $1/x^{3/2}$ prefactor, by arguments similar to those
given for $<S^z_i>$. In Fig.~\ref{fig:endz} we show our results for
$<S^z_1S^z_i>- <S^z_1><S^z_i>$.
As is clearly seen in this figure this correlation function
not only has a staggered part but also a uniform part.
We obtain
\begin{equation}
<S^z_1S^z_i>- <S^z_1><S^z_i>\sim (-1)^{i-1}0.117(1)K_0((i-1)/3.66(1))
-0.081(1)\frac{e^{-2(i-1)/8.33(3)}}{\sqrt{i-1}},
\end{equation}
with $\chi^2=0.31$.
Our data are clearly consistent with both
the $K_0$ term and a uniform term being present in 
$<S^z_1S^z_i>$. Furthermore the value we obtain
for $\xi_\perp$ as well as
for $\xi_{||}$,
\begin{equation}
\xi_{||}=3.66(1),
\end{equation}
is in very good agreement with what we obtain for
the same correlation length from the bulk correlation
function, $<S^z_iS^z_j>$.
However, there are unexplained discrepancies with the predictions of
the free boson model.  A pure exponential alternating piece was not
measured and the uniform part decays with a $1/\sqrt{x}$ prefactor,
rather than $1/x^{3/2}$. We cannot
exclude the possibility that this discrepancy somehow is generated
by the iterative numerical procedure, although we find it unlikely.

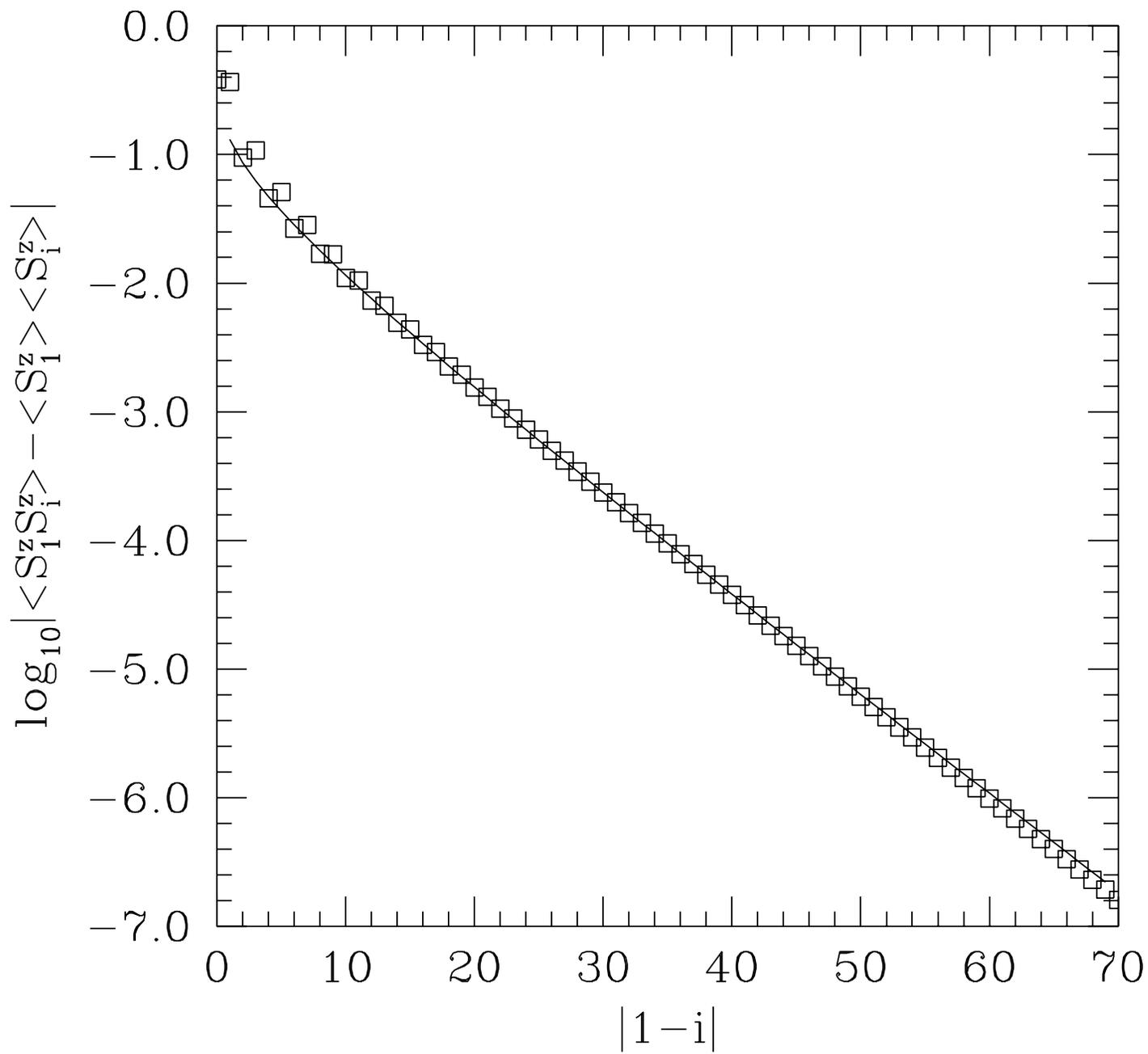
\begin{figure}
\caption{The boundary correlation function, $<S^z_1S^z_i>-<S^z_1><S^z_i>$,
for a 100 site
{\it isotropic} chain. Shown is $<S^z_{1}S^z_i>-<S^z_1><S^z_i>$ as a function
of $|1-i|$. The end of the chain is thus at the left hand
side of the figure.
}
\label{fig:endiso}
\end{figure}

In Fig~\ref{fig:endiso} is shown the boundary correlation function
$<S^z_1S^z_i>-<S^z_1><S^z_i>$ for a 100 site {\it isotropic }
chain. In this case we fit to the simple form
\begin{equation}
<S^z_1S^z_i>- <S^z_1><S^z_i>\sim (-1)^{i-1}0.07(1)K_0((i-1)/5.93(3)),
\end{equation}
with $\chi^2=13.7$.
In this case we do not see an exponential term in the staggered
part nor do we find any uniform part. The obtain correlation
length is in reasonable agreement with what we obtained from the
bulk correlations. In this case we were not able to fit the
uniform part of the correlation function.

\section{Equal Time Structure Factor}\label{sec:struct}
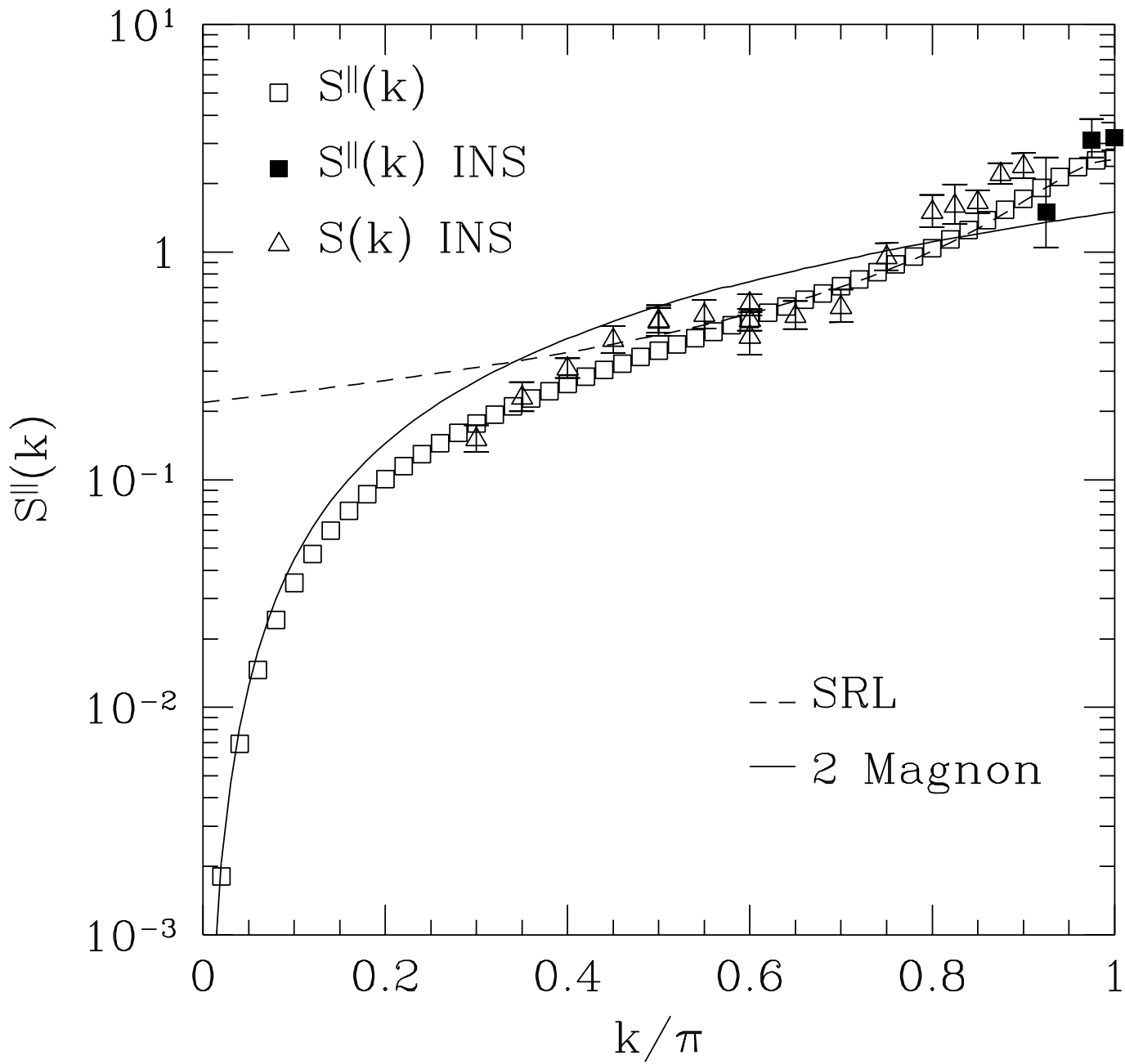
\begin{figure}
\caption{The structure factor, $S^{||}$, as a function of $k/\pi$ for
a 100 site {\it anisotropic} chain. The numerical results are shown
as open squares. INS data from Ref.~\protect\onlinecite{ma}
are shown as solid squares for the data points where
$S^{||}$ experimentally could be resolved
and as open triangles for the data points where to
within experimental accuracy $S^{||}$ and $S^{\perp}$ were
identical. The dashed line is the SRL form obtained
by fitting close to $k=\pi$. The solid line is the prediction
from the free boson theory for the
contribution to $S^{||}$ stemming from two magnon excitations.
}
\label{fig:spar}
\end{figure}

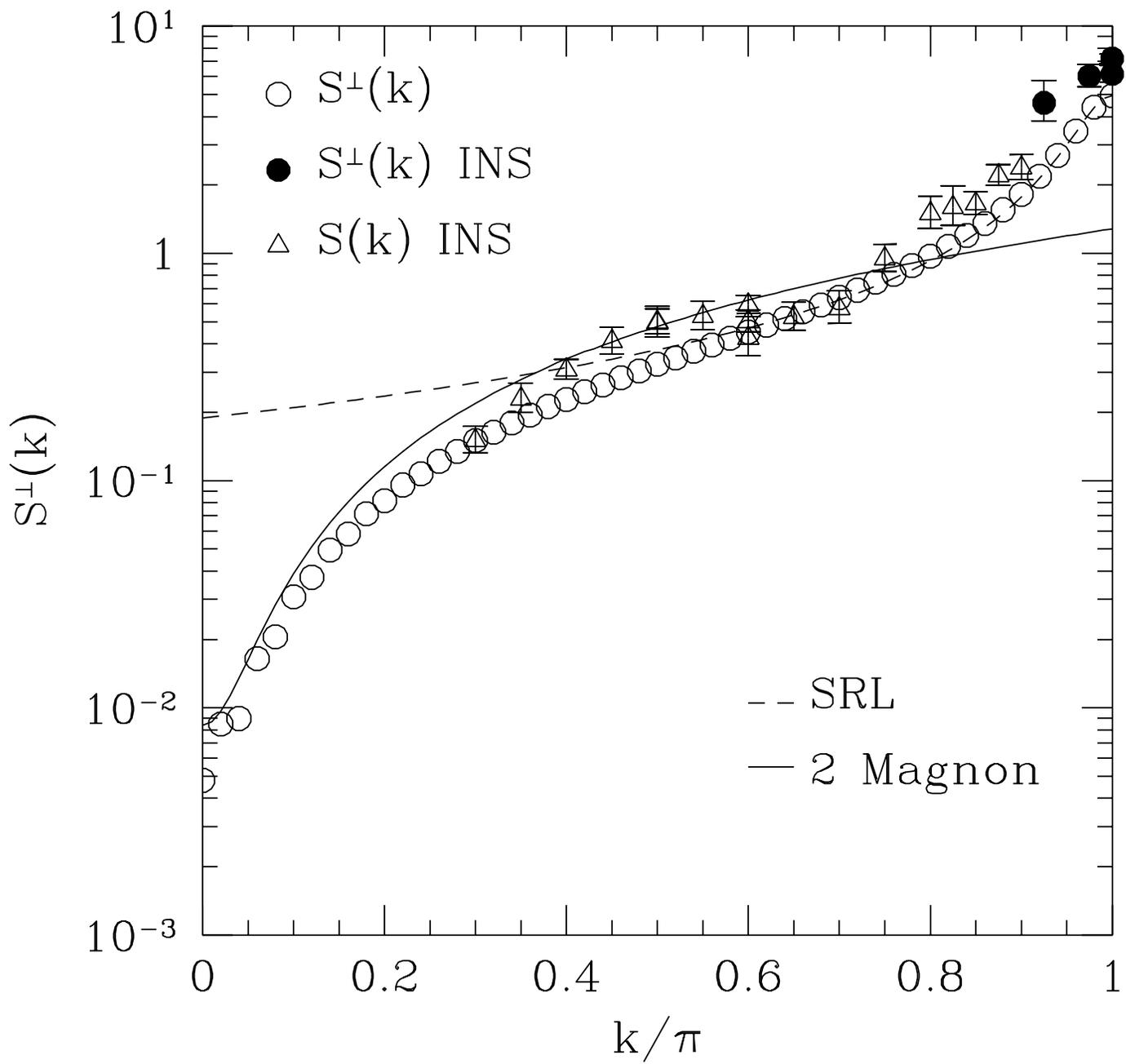
\begin{figure}
\caption{The structure factor, $S^{\perp}$, as a function of $k/\pi$ for
a 100 site {\it anisotropic} chain. The numerical results are shown
as open circles. INS data from Ref.~\protect\onlinecite{ma}
are shown as solid circles for the data points where
$S^{||}$ experimentally could be resolved
and as open triangles for the data points where to
within experimental accuracy $S^{||}$ and $S^{\perp}$ were
identical. The dashed line is the SRL form obtained
by fitting close to $k=\pi$. The solid line is the prediction
from the free boson theory for the
contribution to $S^{||}$ stemming from two magnon excitations.
}
\label{fig:sperp}
\end{figure}

Although, as mentioned, it is not possible to obtain direct
information about dynamical properties using the DMRG
method, it is in principle possible to obtain the equal time 
structure factor,
by simply
Fourier transforming the bulk correlation functions.
Exact diagonalization results for $S(k)$ for shorter chains,
$L=16$, allows us to check the validity of this procedure.
Here we use the standard definition~\cite{lovesey}
\begin{equation}
S(k,\omega)=\frac{1}{2\pi}\sum_r \int dt 
e^{-i\omega t}e^{-ikr}<S(r,t)\cdot S(0,0)>.
\end{equation}
With this definition the equal time structure factor can
be written in the following way
\begin{equation}
S(k)\equiv\int d\omega S(k,\omega)=
\sum_r e^{-ikr}<S(r,0)\cdot S(0,0)>.
\end{equation}
In particular, we use for our data the relation
\begin{equation}
S(k)=<S(50,0)S(50,0)>+2\sum_r e^{-ik(50-r)}<S(r,0)S(50,0)>.
\end{equation}
In Fig.~\ref{fig:spar},~\ref{fig:sperp} this is done so as
to obtain $S^{||}(k)$ and $S^{\perp}(k)$, respectively.
Our results are shown as the open squares and circles, respectively.
The real space bulk correlation functions,
Fig.~\ref{fig:bulkx},~\ref{fig:bulkz},
we have used in obtaining the structure factors were calculated
for an open ended chain and one might wonder about the validity of this
approach. However, since we are able to obtain results
for very long chains the approximation made by assuming
that the correlation functions are equal to the correlation
functions for a {\it periodic} chain with the same length is
very small. For $S^{||}$ the error is negligible since the
associated correlation function decays very fast. However,
for $S^{\perp}$, as can be seen in Fig.~\ref{fig:bulkx}, there
are some end effects present in the correlation function 
which we estimate to give rise to an error of 0.05\% in $S^{\perp}(k=\pi)$,
by using the fitted Bessel function form instead of the
actual data points for $r<20$. Away from $k=\pi$ we expect
the error to be of the same order of magnitude, although somewhat larger
near $k=0$. The advantage
of using the simple Fourier transform is that the full 
weight of the structure factor is conserved. That is the
total moment sum rule,
\begin{equation}
\frac{1}{L}\sum_k [2S^{\perp}(k)+S^{||}(k)]=S(S+1) =2,
\label{eq:srule}
\end{equation}
is obeyed to within numerical precision.

The dotted lines in Fig.~\ref{fig:spar},~\ref{fig:sperp} are
square root Lorentzians (SRL). We define this
as
\begin{equation}
g_a\frac{\xi_a}{2}\frac{1}{\sqrt{1+(k-\pi)^2(\xi_a)^2}},\ \
\xi_a=v_a/\Delta_a,
\label{eq:srl}
\end{equation}
which is just the Fourier transform of the Bessel function form
of the staggered part of the correlation function, Eq.~(\ref{eq:stbes}).
In both figures we have used the values previously obtained
for $g_a$  and $\xi_a$. From section~\ref{sec:bulkcor} we know that the
SRL form should work very well close to $k=\pi$, since $g_a, \xi_a$ was
obtained by fitting to the staggered part of the correlation functions.
From Fig.~\ref{fig:spar},~\ref{fig:sperp} it is, however, apparent that the
simple
SRL form only describes the structure factor well from $k/\pi\sim 0.7$
to $k/\pi=1$.

For the isotropic chain the ground-state is a singlet and
therefore $\sum_i S(i,t)|0>=0$. Thus the structure factor
must vanish as $k^2$ close to $k=0$. If we introduce anisotropy
then $S^{||}$ still vanishes as $k^2$ close to $k=0$ but
$S^\perp$ can take on a non zero value at $k=0$~\cite{aw}.
Within the frame work of the free boson approximation it
is possible to derive approximate expressions for the structure
factors in the presence of anisotropy. This has previously been
done assuming uniform velocities $v=v_x=v_y=v_z$ in which case
simple expressions for $S(k,\omega)$ can be derived~\cite{aw}.
We now generalize these results to allow for different
velocities but focus only on the equal time structure factor, $S(k)$,
since expressions for the full dynamical structure factor, $S(k,\omega)$,
in this case will be rather complicated. 
The structure factor close to $k=\pi$ will be completely
dominated by correlations in the uniform part of the magnetization
density. If we use the general expression
for ${\bf l}$, Eq.~(\ref{eq:l}), along with the mode expansion,
Eq.~(\ref{eq:modexp}), and the commutation relations, Eq.~(\ref{eq:comrel}),
we find that ${\bf l}$ will contain a double creation term of the following
form for $l^x$ (compare Eq.~(\ref{eq:lc}))
\begin{equation}
l^{x}_c(x,t) = i\int\frac{dk_ydk_z}{16\pi^2\omega_{k_y}\omega_{k_z}}
e^{i(K_y+K_z)\cdot X}(\frac{\omega_{k_z}}{v_z}-\frac{\omega_{k_y}}{v_y})
a^{\dag}_{k_y}a^{\dag}_{k_z}.
\end{equation}
Again this term will be the only term contributing to
$<l^x(x_i)l^x(x_j)>$ in the ground state. We thus obtain for
the equal time correlation function,
\begin{equation}
<0|l^x(x,0)l^{x}(0,0)|0>=
\int\frac{dk_ydk_z}{16\pi^2\omega_{k_y}\omega_{k_z}}
e^{i(k_y+k_z)x}(\frac{\omega_{k_z}}{v_z}-\frac{\omega_{k_y}}{v_y})^2v_yv_z.
\end{equation}
Thus we see that the two magnon contribution to $S^{\perp}(k)\equiv
S^{xx}(k)=S^{yy}(k)$ is
\begin{equation}
a^2\int \frac{dk_ydk_z}{8\pi}(\frac{\omega_{k_z}v_y}{\omega_{k_y}v_z}+
\frac{\omega_{k_y}v_z}{\omega_{k_z}v_y}-2)\delta(k-k_y-k_z).
\label{eq:sperp}
\end{equation}
For the parallel mode the two magnon contribution to
the structure factor, $S^{||}(k)\equiv S^{zz}(k)$, becomes simply
\begin{equation}
a^2\int \frac{dk_xdk_y}{8\pi}(\frac{\omega_{k_x}}{\omega_{k_y}}+
\frac{\omega_{k_y}}{\omega_{k_x}}-2)\delta(k-k_x-k_y),
\label{eq:spar}
\end{equation}
which is just the Fourier transform of Eq.~(\ref{eq:ll}).
Here the overall scale is {\it not} a free parameter, as opposed
to Eq.~(\ref{eq:srl}) where $g_a$ is a free parameter,
because $\int dx {\bf l}=\sum {\bf S}_i$ must obey spin commutation
relations.
These two predictions are shown as the solid lines in
Fig.~\ref{fig:spar},~\ref{fig:sperp}. Qualitatively the free
boson estimates seems to agree well with the
numerical results for $k$ close to 0. For $k/\pi\geq 0.1$
the free boson estimate for the two magnon
part of $S(k)$ is larger than the numerical results for
the full structure factor. Since the free boson theory
doesn't take interaction effects into account this
is perhaps not too surprising.

In Fig.~\ref{fig:spar},~\ref{fig:sperp} are also shown the 
inelastic neutron scattering (INS) results of Ref.~\onlinecite{ma}
which are directly comparable to our results. 
The open triangles are points where to within experimental accuracy
$S^{||}$ and $S^{\perp}$ were identical. The full squares and
circles are data points where $S^{||}$ and $S^\perp$, respectively,
could be resolved experimentally. 
Good agreement is  obtained with the experiment apart from
an overall scale factor of about 1.25 which was to be expected since the
experimental total intensity exceeded the exact sum rule:
$(1/L)\sum_{\alpha
\tilde k}S^{\alpha \alpha}(\tilde k) = s(s+1)$ by 30\% ($\pm$ 30\%).
A very nice agreement 
between the numerical results and experimental data is evident.

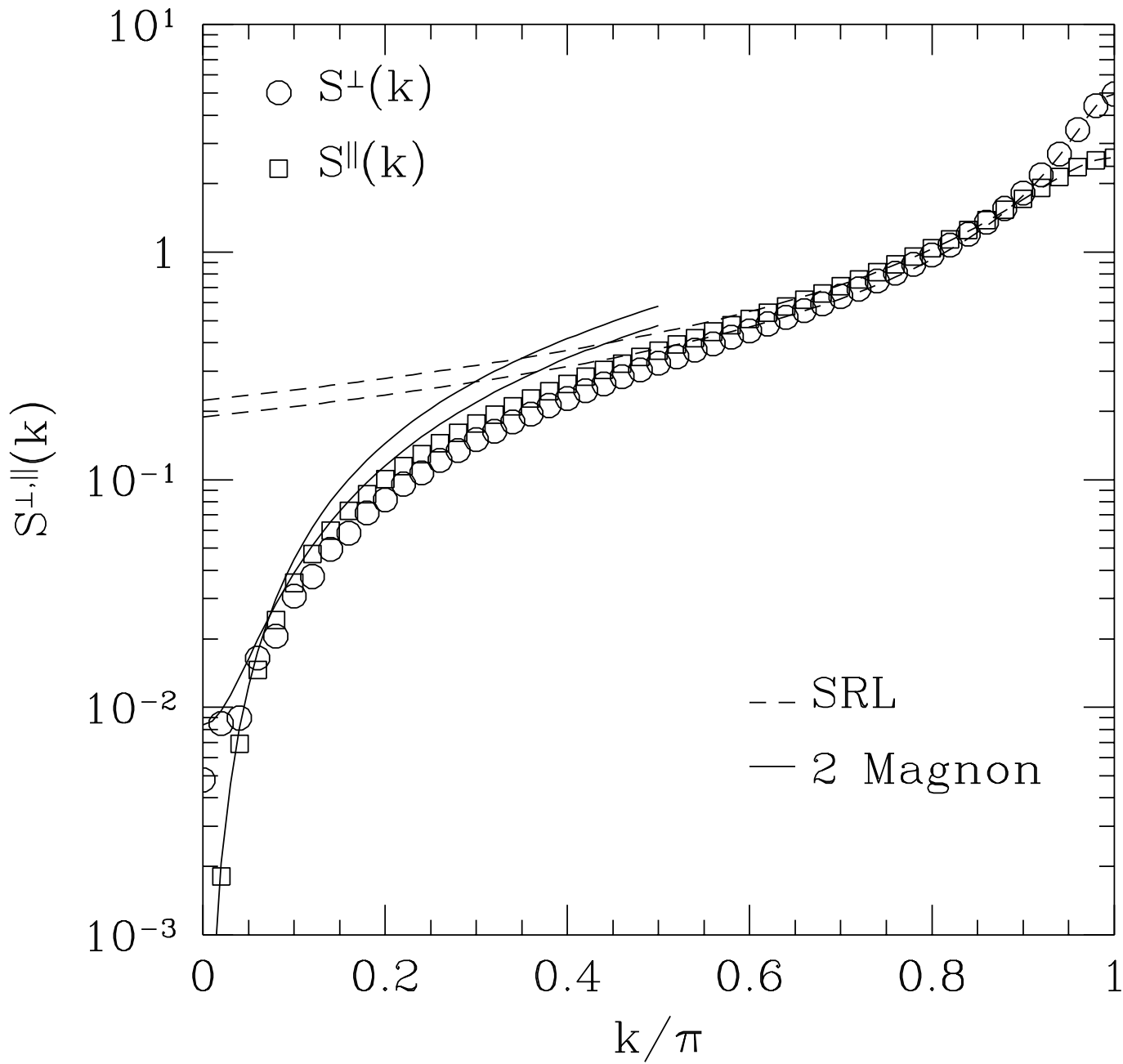
\begin{figure}
\caption{The structure factors, $S^{||}$ and $S^{\perp}$ 
as a function of $k/\pi$ for
a 100 site {\it anisotropic} chain. The numerical results are shown
as open squares and circles, respectively.
The dashed lines are the SRL form obtained
by fitting close to $k=\pi$. The solid lines 
are the predictions from the free boson theory for the
contribution to the structure factor stemming from two magnon excitations.
}
\label{fig:sboth}
\end{figure}

Fig.~\ref{fig:sboth} summarizes our results for
$S^{||}$ and $S^\perp$. In this figure the two structure factors
are plotted together with the different theoretical
estimates. For the range $(0.1-0.85)k/\pi$ $S^{||}$ is the
larger of the two, and only when $k/\pi\geq 0.85$ or $k/\pi<0.1$
does
$S^{\perp}$ become dominant. The crossing of the two structure
factors near $k/\pi\sim0.85$
is correctly described by the SRL if
different velocities for the two modes are allowed for. The
crossing near $k/\pi\sim 0.1$ seems also to be predicted by
the free boson theory estimate for the two magnon part
of the structure factor. This seems also to be supported by
exact diagonalization studies for $L=16$~\cite{golinelli3}.

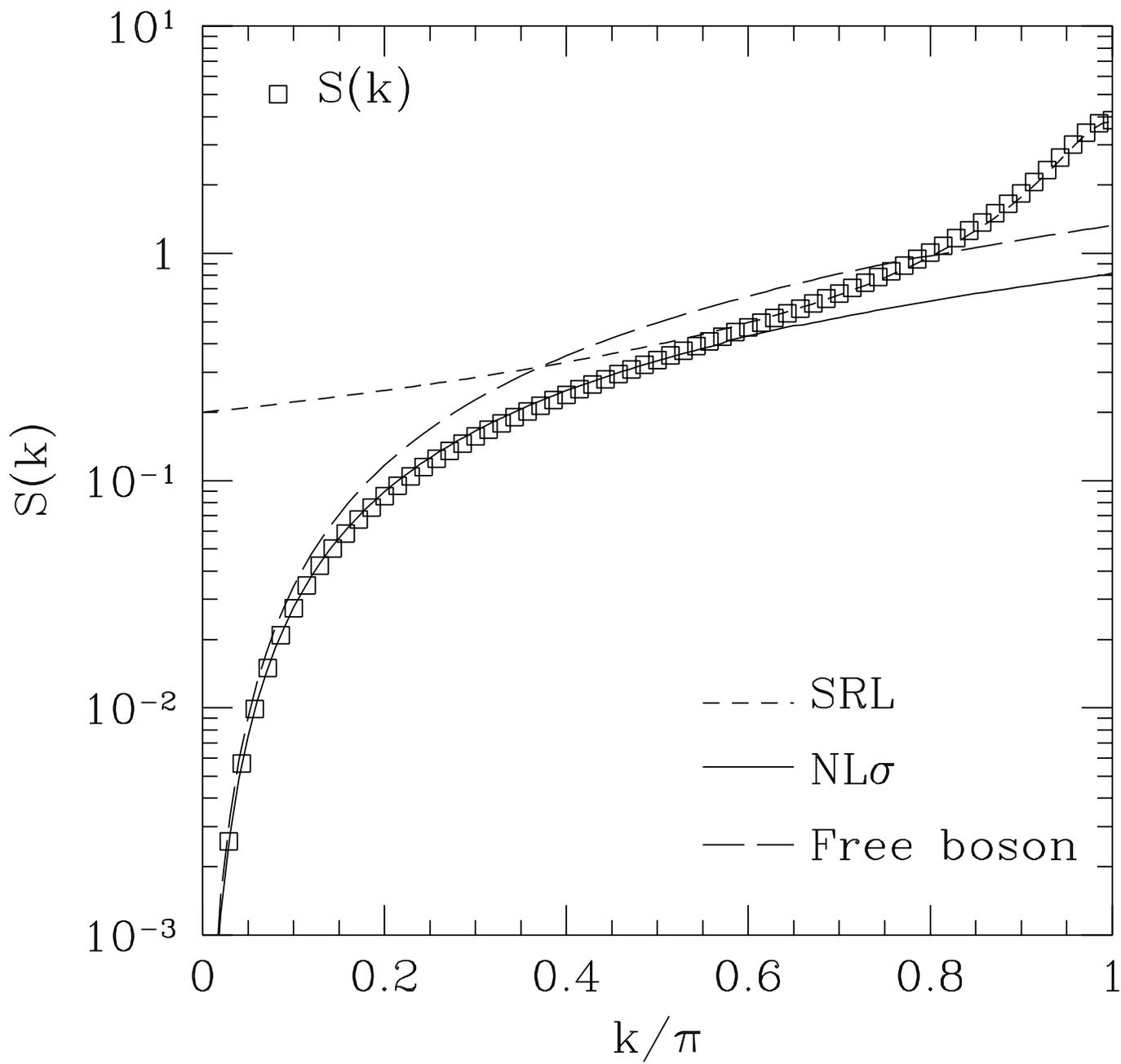
\begin{figure}
\caption{The structure factor, $S$, as a function of $k/\pi$ for
a 100 site {\it isotropic} chain. The numerical results are shown
as open squares. 
The short dashed line is the SRL form obtained
by fitting close to $k=\pi$. The solid line 
is the exact prediction from the NL$\sigma$ model
for the
contribution to the structure factor from two magnon excitations.
The long dashed line is the free boson prediction for the 
two magnon contribution to the structure factor.
}
\label{fig:siso}
\end{figure}

In the case of the {\it isotropic} chain, with $D=0.0$,
it is possible to obtain an exact expression for the
two magnon part of $S(k\omega)$ within the frame work
of the NL$\sigma$ model~\cite{aw} thereby taking into account interaction
effects. 
In Ref.~\onlinecite{aw} it is shown that 
with $S=S^x=S^y=S^z$, the two magnon contribution to $S$ is given by
\begin{equation}
|G(\theta)|^2\frac{vk^2}{2\pi}\frac{\sqrt{{\bf K}\cdot{\bf K}
-4\Delta^2}}{({\bf K}\cdot{\bf K})^{3/2}},\ \ {\bf K}\cdot{\bf K}>4\Delta^2.
\label{eq:skwnl}
\end{equation}
Here ${\bf K}\cdot{\bf K}=4\Delta^2{\rm cosh}^2(\theta/2)$ and $|G(\theta)|^2$
is given by the expression
\begin{equation}
|G(\theta)|^2=\frac{\pi^4}{64}\frac{1+(\theta/\pi)^2}{1+(\theta/2\pi)^2}
\left(\frac{{\rm tanh}(\theta/2)}{\theta/2}\right)^2.
\end{equation}
This calculation contains no free parameters.
The expression for the two magnon contribution to $S(k,\omega)$
can now easily be integrated numerically over
$\omega$ to obtain $S(k)$. 
Our numerical results
are shown in Fig.~\ref{fig:siso} along with the SRL form using the
previously obtained values for $v, g$ and $\xi$. The SRL form
is shown as the short dashed line. Also
shown, as the solid line, is the NL$\sigma$ model
results for the two magnon part of $S(k)$. 
As seen in Fig.~\ref{fig:siso} there is
excellent agreement between the theoretical
and numerical results. The long dashed line is the 
prediction from the free boson theory for the two magnon contribution
to the structure factor. The inclusion of interaction effects changes
the shape of $S(k)$ and we see that the free boson estimate, although
qualitatively correct near $k=0$, somewhat overestimates $S(k)$ for
larger $k$. Presumably the NL$\sigma$ model also
predicts a 4-magnon (and higher) contribution
to $S(k,\omega)$. This is not
known exactly and not included in Eq.~(\ref{eq:skwnl}). Hence the full
$S(k)$ in the NL$\sigma$ model should be somewhat larger than
Eq.~(\ref{eq:skwnl}). The very precise agreement with the numerical
results may indicate
that this multi-magnon contribution is very small. Alternatively, it may
indicate that the NL$\sigma$ model somewhat overestimates the two-magnon part.
Our numerical results for $S(k)$ are also in good agreement with
previous results using Monte Carlo techniques~\cite{nomura,liang}
for chains of length 64.

\section{Single Mode Approximation}\label{sec:sma}
As already mentioned the DMRG method does not allow us to
obtain information about the dynamical properties of the system.
However, if $S(k,\omega)$ is dominated by an intense single mode,
as is the case for $k$ close to $\pi$, then we can
obtain an approximate dispersion relation through the use of
the single mode approximation (SMA). This has previously been done
for the isotropic chain~\cite{tak88}. This can then be compared to
exact diagonalization results~\cite{golinelli3,tak93,haas} and to
QMC results for the isotropic chain~\cite{tak1,meshkov1}.
We assume that for both modes
\begin{equation}
S(k,\omega)=S_0(k)\delta(\omega-\omega_k)+\tilde{S}(k,\omega),
\label{eq:sma}
\end{equation}
where $\tilde{S}(k,\omega)$ is non zero only for $\omega> \omega_c$
and $\omega_c > \omega_k$.
Here $\omega_c$ denotes the bottom of the {\it continuous} part
of the spectrum.
We then look at the integral
\begin{equation}
\frac{1}{\omega_k}\int_{-\infty}^{\infty}d\omega \omega S(k,\omega)=S_0(k)+
\frac{1}{\omega_k}\int_{-\infty}^{\infty}d\omega \omega\tilde{S}(k,\omega).
\end{equation}
Since, by assumption, $\tilde{S}(k,\omega)$ only is non zero
when $\omega > \omega_c > \omega_k$ we must have
\begin{equation}
\frac{1}{\omega_k}\int_{-\infty}^{\infty}d\omega \omega\tilde{S}(k,\omega)\geq
\int_{-\infty}^{\infty}d\omega \tilde{S}(k,\omega).
\end{equation}
We can then write
\begin{equation}
\frac{1}{\omega_k}\int_{-\infty}^{\infty}d\omega \omega S(k,\omega)\geq
S_0(k)+\int_{-\infty}^{\infty}d\omega \tilde{S}(k,\omega)=S(k),
\label{eq:smat}
\end{equation}
where $S(k)$, per definition, is 
the equal time structure factor. The first moment of the dynamical 
structure factor
obeys  a simple sum rule~\cite{lovesey,hohenberg} 
\begin{equation}
\int_{-\infty}^{\infty} d \omega \omega S^{xx}(k,\omega)=
-\frac{1}{2}<[[H,S^x(k)],S^x(-k)]>,
\end{equation}
with equivalent relations for $S^{yy}(k,\omega)$ and $S^{zz}(k,\omega)$.
Here we have used
\begin{equation}
S^a(k)=L^{-1/2}\sum_i e^{-ikx_i}S^a_i.
\end{equation}
Explicitly
calculating the double commutator we get 
the following relation
\begin{eqnarray}
\int_{-\infty}^{\infty}d\omega \omega S^{xx}(k,\omega)&=&
-J[F_y-F_z\cos(k)+F_z-F_y\cos(k)+D(G_z-G_y)]\nonumber\\
\int_{-\infty}^{\infty}d\omega \omega S^{zz}(k,\omega)&=&
-J[F_x-F_y\cos(k)+F_y-F_x\cos(k)],
\end{eqnarray}
where $F_a=<S^a_iS^a_{i+1}>$ and $G_a=<(S^a_i)^2>$.
From Eq.~(\ref{eq:smat}) we then get the SMA dispersion 
relations
\begin{eqnarray}
\omega^\perp_k&\leq&\omega^{\perp}_{SMA}(k)
=-\{J(F_y+F_z)(1-\cos(k))+D(G_z-G_y)\}/S^\perp(k)\nonumber\\
\omega^{||}_k&\leq&\omega^{||}_{SMA}(k)
=-\{J(F_x+F_y)(1-\cos(k))\}/S^{||}(k).
\label{eq:wsma}
\end{eqnarray}
We see that we always have $\omega_{SMA}\geq\omega_k$, where
$\omega_k$ describes the dispersion of the singular mode.
The inequality is {\it only} fulfilled as
long as $\omega_k$ is {\it below} the
edge of the continuous part of the spectrum, $\omega_c$,
see Eq.~(\ref{eq:sma}).
If we denote the lower edge of the spectrum by $\omega_e$, then
it is easy to see~\cite{tak88} that the inequality
$\omega_{SMA}\geq\omega_e$ always is satisfied independent of the position
of any singular modes. This simply follows from the trivial inequality
\begin{equation}
\frac{1}{\omega_e}\int_{-\omega_e}^{\infty}d \omega \omega S(k,\omega)
\geq \int_{-\omega_e}^{\infty}d \omega S(k,\omega) = S(k).
\end{equation}
Retracing the above steps we find $\omega_{SMA}\geq\omega_e$.

The long-distance field theory limit of the antiferromagnetic
Heisenberg model is
the $O(3)$
NL$\sigma$ model, Eq~(\ref{eq:lnl}),
describing a massive triplet of fields.
The unit vector $\bbox{\phi}$ of fields describes the staggered
magnetization of the spin chain.
The spectrum for the $S=1$ spin chain
at $k=\pi$ should therefore be dominated by large single modes
corresponding to the massive triplet together with a small continuum
from
3,5,\ldots magnons. The NL$\sigma$
model has no bound states and the spectrum at $k=0$
must therefore be a continuum corresponding to 
the excitation of two massive particles with $k=\pm\pi$ together with
a small contribution from 4,6,\ldots magnons.
If anisotropy is present the spectrum at $k=\pi$ consists of a
light magnon with mass $\Delta_\perp$ and $S^z_T=\pm 1$ and a heavy magnon
with $S^Z_T=0$ and mass $\Delta_{||}$.
At $k=0$ the lower edge of the continuum of two magnon
excitations for out of plane modes,
$\omega^{||}$, which have $S^z_T=0$ with respect to the ground-state,
must be given by the excitation of two
light magnons with $k=\pm\pi$ and $S^z_T=\pm 1$, corresponding
to a gap of $2\Delta_\perp$. For in-plane modes,
$\omega^{\perp}$,
which have $S^z_T=\pm 1$ with respect to the ground-state, the lower
edge of the two magnon continuum at $k=0$ must correspond to one
light magnon with $S^z_T=\pm 1$ and one heavy magnon $S^z_T=0$ and
opposite momentum. Thus the gap should in this case be $\Delta_\perp
+\Delta_{||}$. One expects that somewhere between $k=0$ and
$k=\pi$ the lowest energy excitations will change from the single mode
excitations at $k=\pi$ to the two magnon continuum close to $k=0$.
This picture of the spectrum at $k=0$ is in
good agreement with exact diagonalization results~\cite{golinelli3}.

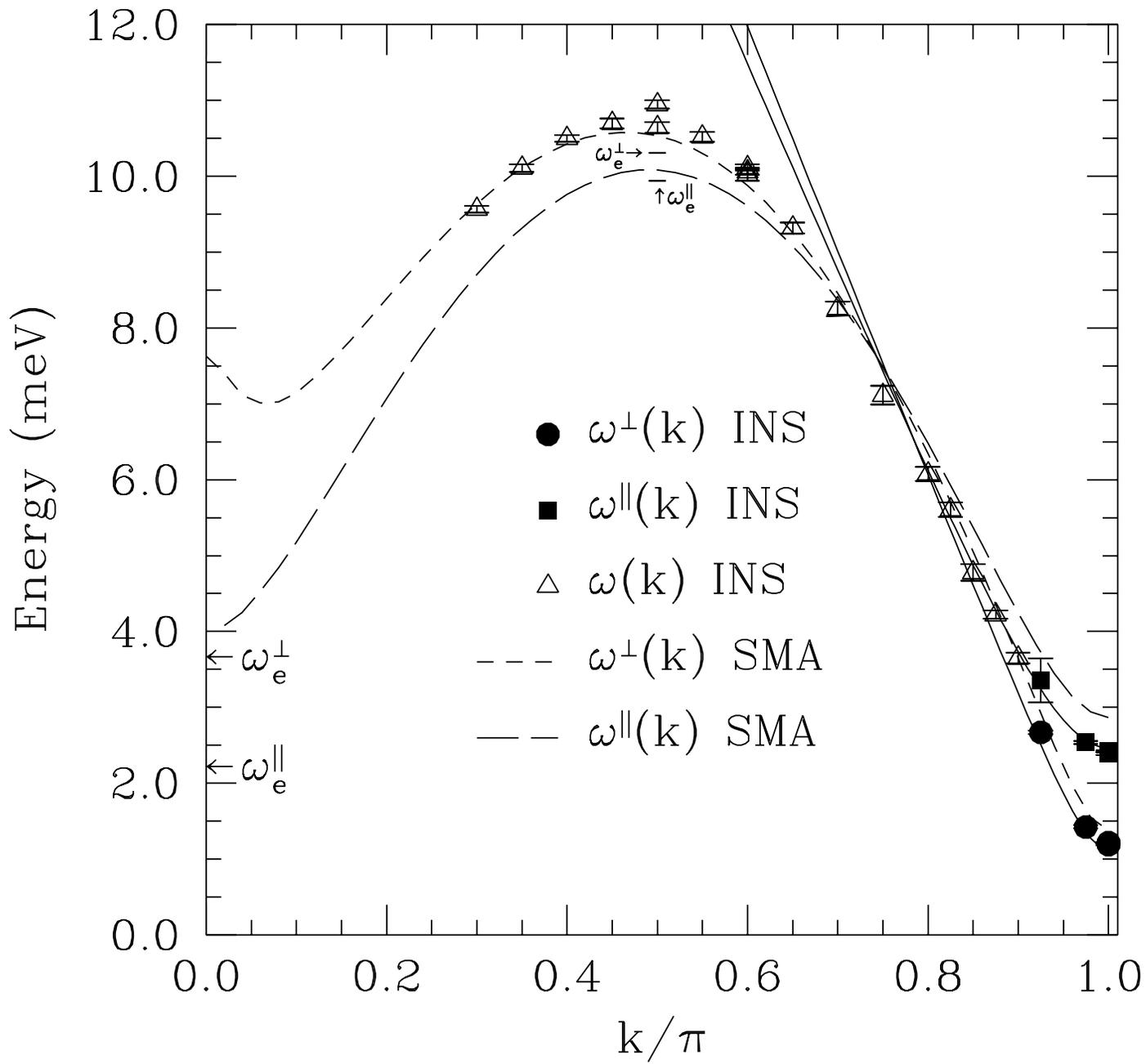
\begin{figure}
\caption{Dispersion of the excitations 
as a function of $k/\pi$ for
a 100 site {\it anisotropic} chain.
The points are INS data from Ref.~\protect\onlinecite{ma}.
The solid lines are the dispersion relations
$\omega^{||,\perp} (k) = J\protect\sqrt{\Delta^2_{||,\perp}+v^2_{||,\perp}(k-\pi)^2}$,
for the two modes valid close to $k=\pi$.
A value of $J=3.75$meV has been used to scale the data with
respect to the experimental results.
The long dashed line is $\omega^{||}_{SMA}$,
and the short dashed line is $\omega^{\perp}_{SMA}$.
The four arrows indicate the lower edge of the excitation spectrum at
$k=0$, $\omega_e^{\perp}=0.986J$ and $\omega_e^{||}=0.60J$, and
at $k=\pi/2$, $\omega_e^{\perp}=2.75J$ and $\omega_e^{||}=2.65J$,
from
the exact diagonalization results of
Ref.~\protect\onlinecite{golinelli3}.
}
\label{fig:omanis}
\end{figure}

For the anisotropic
Heisenberg chain, with $D/J=0.18$, we have
$F_x=F_y=-0.4969462$ and $F_z=-0.4035174$. In conjunction with
$<(S^z_{50})^2>=0.6206864$ this agrees with the previously obtained
value
for $e_0/J= -1.2856861$ to within numerical precision. Using
Eq.~(\ref{eq:wsma}) along with $<(S^x_{50})^2>=0.6896568$
we can now obtain the dispersion relations within
the SMA. Our results are shown as the dashed lines
in Fig.~\ref{fig:omanis} along with the
experimentally determined dispersion relation from
Ref.~\onlinecite{ma}. 
The long dashed line is $\omega^{||}_{SMA}$,
and the short dashed line is $\omega^{\perp}_{SMA}$.
The INS data from Ref.~\protect\onlinecite{ma}
are shown as solid squares and circles for the data points where
$\omega^{||}$ and $\omega^\perp$, respectively, could be resolved
and as open triangles for the data points where to
within experimental accuracy $\omega^{||}$ and $\omega^{\perp}$ were
identical. The two previously determined velocity, $v_{||}$ and
$v_{\perp}$, gives us  dispersion relations close to $\pi$ which
are shown as solid lines. 
Also indicated in Fig.~\ref{fig:omanis} is the lower edge
of the spectrum at $k=0$, $\omega_e^{\perp}=0.986J$ and 
$\omega_e^{||}=0.60J$, and at $k=\pi/2$,
$\omega_e^{\perp}=2.75$ and $\omega_e^{||}=2.65J$,
obtained from the exact diagonalization of Golinelli at
al~\cite{golinelli3}.
The numerical data is scaled with
$J=3.75$ meV. The agreement between the experimental results
and our dispersion relations (solid lines) close to $k=\pi$ is excellent.
The dispersion relations are also in very good agreement with the
experimental results of Ref.~\onlinecite{renard87}.
The solid lines indicate a crossing of $\omega^{||}$ and $\omega^\perp$
at approximately $k/\pi\sim0.8$. This has also been observed in exact
diagonalization studies~\cite{golinelli3} and QMC studies~\cite{deisz2}
and seems to agree with
the SMA prediction as well as with the fact that experimentally
$\omega^\perp$ and $\omega^{||}$ only can be resolved for
$k/\pi>0.9$. As expected we see that $\omega_{SMA}$ consistently
is {\it larger} than the experimental results and our relativistic
dispersion relations when $k$ is close to $\pi$. Thus there is 
clearly a small but non zero 
contribution to $S(k,\omega)$ from multi magnon processes
at $k\sim\pi$. $\omega_{SMA}^{||}$ becomes
smaller than the experimental points at around $k/\pi\sim0.7$. 
However, $\omega_{SMA}^{\perp}$ coincides with the experimental data
all the way to $k/\pi=0.3$. 
One should note
that a complete agreement between $\omega_{SMA}$ and the experimental
data doesn't necessarily mean that the single mode approximation
is exact but can be due to the fact that the singular mode is no longer
below the continuous part of the spectrum.
At $k/\pi=0.5$ both SMA estimates are
slightly above the lower edges $\omega^{\perp,||}_e$ obtained from
exact diagonalization~\cite{golinelli3}. 
The experimental and numerical 
data seem consistent with a well resolved single mode all the
way to $k/\pi=0.3$, but this
may no longer be well separated from a small multi magnon continuum.
The fact that our results for $\omega_{SMA}$ remains rather close
to the experimental data is consistent with a large singular
contribution to $S(k,\omega)$ in this region.
At $k=0$ we can only compare to the exact diagonalization results
of Ref.~\onlinecite{golinelli3}. The SMA results are now at almost twice
the value of these results consistent with a large contribution
from multi magnon processes. It therefore seems likely that
the continuous part of the spectrum will only set in
at $k/\pi<0.3$ in NENP. This is not inconsistent with our findings for
the structure factor.

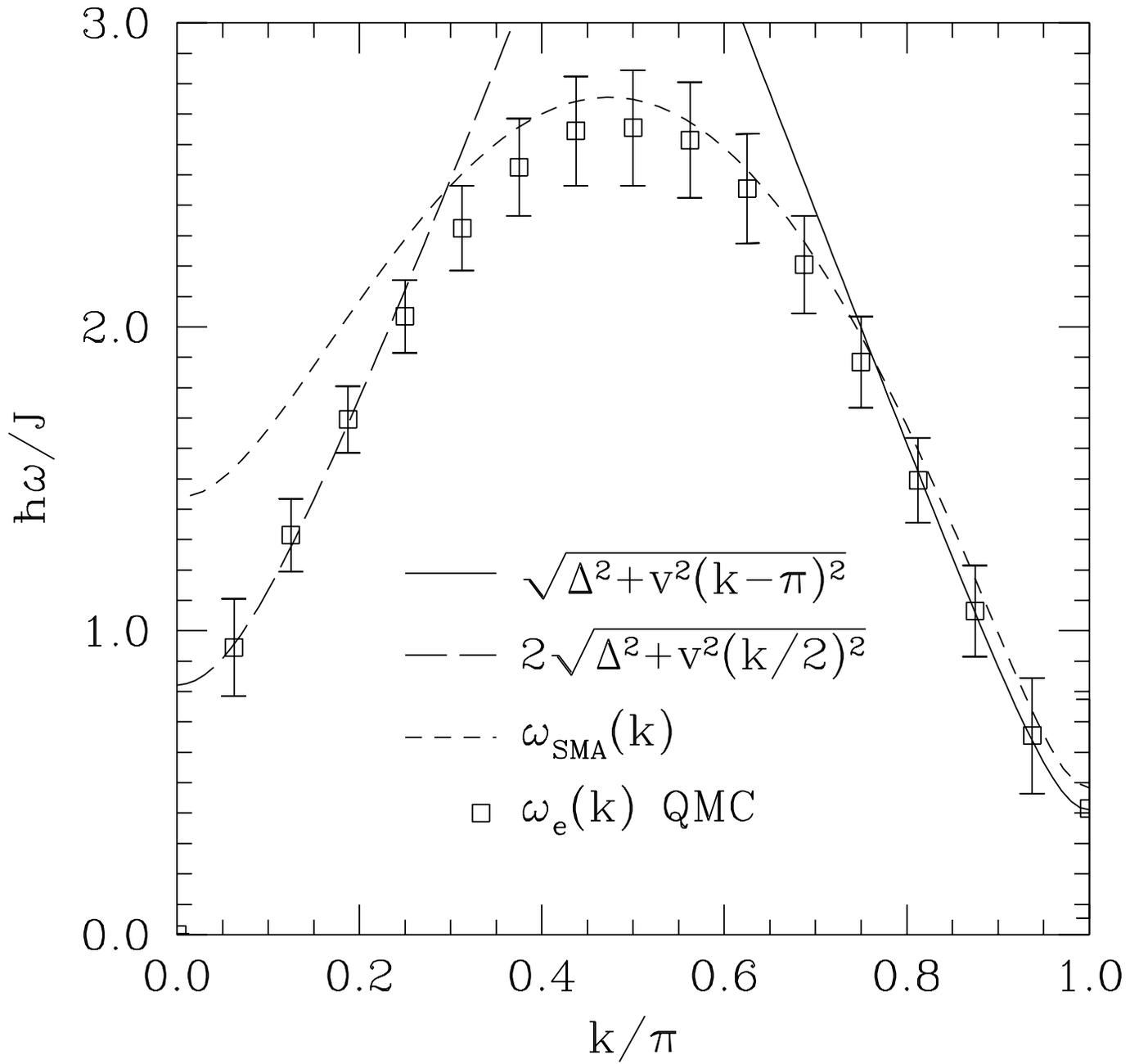
\begin{figure}
\caption{Dispersion of excitations 
as a function of $k/\pi$ for
a 100 site {\it isotropic} chain.
The points are QMC data from Ref.~\protect\onlinecite{tak1}.
The solid line is the dispersion relation
$\omega (k)/J = \protect\sqrt{\Delta^2+v^2(k-\pi)^2}$,
valid close to $k=\pi$. The long dashed line
is the bottom of the
two magnon continuum, $2\protect\sqrt{\Delta^2+v^2(k/2)^2}$, valid near $k=0$.
The dashed line is the SMA prediction for the
dispersion relations obtained from the equal time 
structure factor. The poor agreement between the SMA prediction and the
QMC results at small $k$ indicates the two magnon nature of the
excitations there.
}
\label{fig:omiso}
\end{figure}

For the {\it isotropic} chain we have $F_x=F_y=F_z=e_0/3=-0.4671613$ 
again in very good agreement with the known~\cite{white2} value of
$e_0/J= -1.401484038971(4)$. In Fig.~\ref{fig:omiso} the SMA results are
shown as the dashed line. Also shown in Fig.~\ref{fig:omiso},
as open squares,
are the QMC results of Ref.~\onlinecite{tak1}. The QMC results
should determine the lower edge of the excitations spectrum regardless
of whether this is a single mode or a continuum. The
previously determined velocities and gap~\cite{us1} gives a
relativistic dispersion relation close to $k\sim\pi$ which is
shown as the solid line. The agreement between the relativistic
dispersion relation and the QMC results close to $k\sim\pi$ is
excellent as one would expect. The SMA results indicate again a
small but nonzero contribution to $S(k,\omega)$ from
the multi magnon continuum close to $k=\pi$. For a large region
around $k=\pi$ the SMA seems to work well as compared
to the QMC results. However, close to $k=0$ the SMA is considerably
larger than the QMC results indicating a large contribution to
$S(k,\omega)$ from multi magnon processes.
Also shown in Fig.~\ref{fig:omiso} is the bottom of the two magnon
continuum, $2\protect\sqrt{\Delta^2+v^2(k/2)^2}$, valid near $k=0$.
This result is in very good agreement with the QMC results.

\section{Discussion}
Using the DMRG method it is difficult to get a good handle on
what the actual error-bars are. Presumably the finite lattice method
that most of our results have been obtained with effectively works
as a variational method~\cite{white1,white2}.
The error-bars quoted in the previous
sections are therefore only statistical error-bars obtained by
estimating the errors occurring in the DMRG. This does not include
any systematical errors that are present. It is also not a trivial
matter to fit data obtained on finite lattices to theoretical forms
that are only asymptotically valid. We therefore here recapitulate
our main findings with error-bars we believe are close to reality
but more or less phenomenologically obtained by judging the variation
of the various parameters between different fits.
\begin{eqnarray}
\Delta_{||}/J=0.6565(5),  \ \ &v_{||}/J=2.38(1),\ \ &\xi_{||}/a=3.69(5),\ \ g_{||}\simeq 1.37,\nonumber\ \ \lambda_{||}\simeq 0.85\\
\Delta_\perp/J=0.2998(1), \ \ &v_\perp/J=2.53(1),\ \ &\xi_\perp/a=8.35(7),\ \ 
g_{\perp}\simeq 1.16,\ \ \lambda_{\perp}\simeq 0.53.
\end{eqnarray}
Here all the results were obtained for an anisotropic chain with
$D/J=0.18$. The bulk correlation functions show evidence of
a small but non zero uniform part.
Close to $k=\pi$
the equal time structure factor is very well described by
a square root lorentzian.
The equal time structure
factor $S^{\perp}$ does in this case no longer approach 0 as
$k\rightarrow 0$. In this region both $S^{||}$ and $S^{\perp}$
are well described by
a two magnon contribution to the structure factor
calculated within the free boson theory. The agreement
between the free boson theory and the numerical results is
however only qualitative and judging from the comparison between
a single mode approximation and the experimental data it seems
likely that a single mode persist all the way to $k/\pi=0.3$.
By comparing the single mode approximation
to exact diagonalization results it is seen that
between $k=0$ and $k/\pi=0.3$ 
the continuous part of the spectrum
should become dominant.

For the isotropic chain, with $D/J=0.0$, we find~\cite{us1}
\begin{equation}
\Delta/J=0.4107(1),  \ \ v/J=2.49(1),\ \ \xi_{||}/a=6.03(1),\ \
g\simeq 1.26,\ \ \lambda\simeq 0.72.
\end{equation}
Also in this case we find evidence for a small but non zero
uniform part of the bulk correlation functions. 
Close to $k=\pi$
the equal time structure factor is very well described by
a square root lorentzian.
In the region close to $k=0$ the equal time structure
factor can be compared to an exact result obtained from the
NL$\sigma$ model for the two magnon contribution to
the structure factor. An excellent agreement between theory
and the numerical data is seen out to fairly large values of
$k$. This seems consistent with the behavior of the 
spectrum. Here the single mode approximation as compared
to quantum Monte Carlo results indicates the
onset of a continuous part earlier than for the anisotropic 
chain.

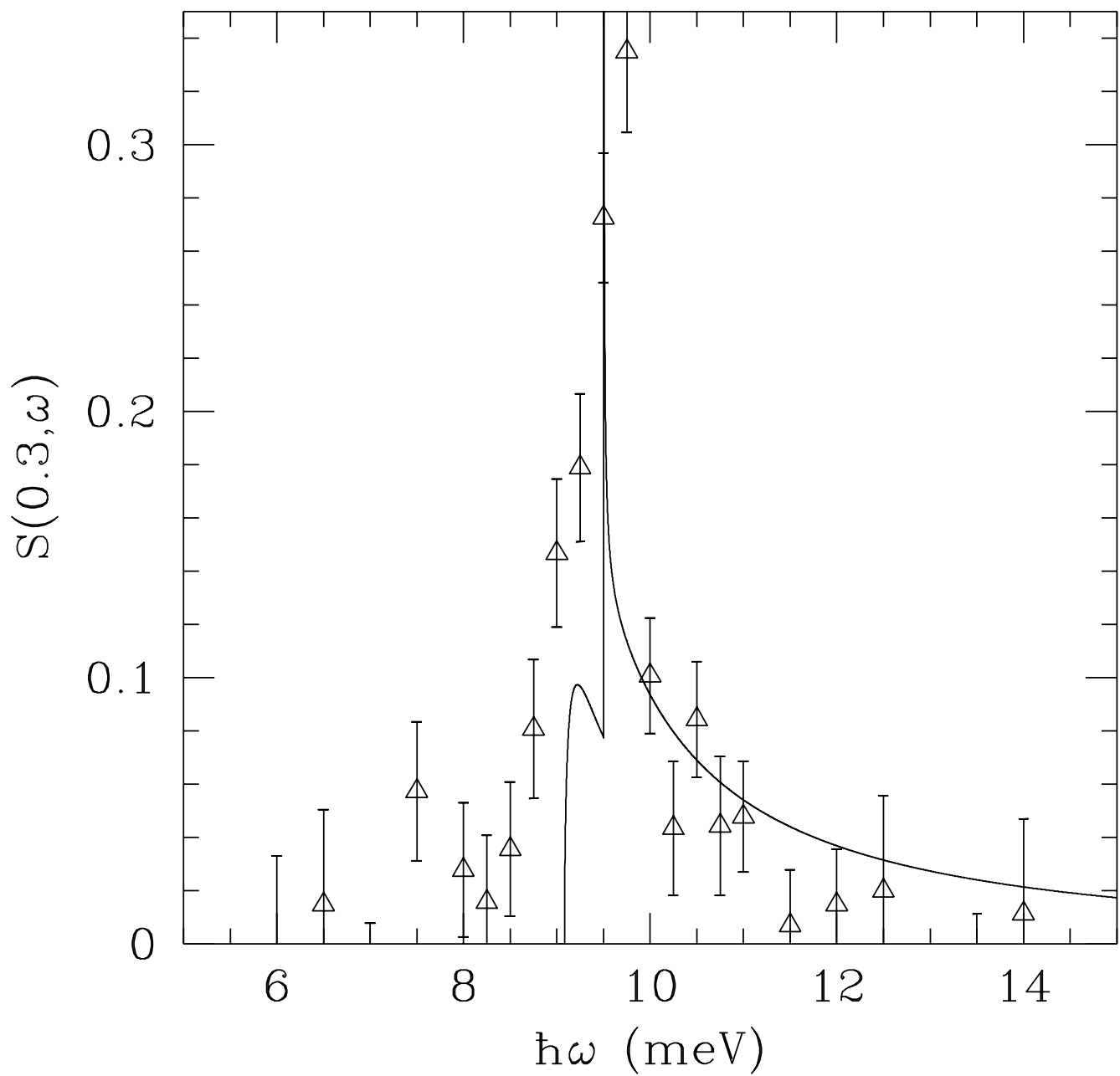
\begin{figure}
\caption{$S(k=0.3\protect\pi,\protect\omega)$ as a
function of $\protect\omega$ for the
free boson model (solid line) and INS results (open triangles) from
Ref.~\protect\onlinecite{ma}. We have used $J=3.75$meV.
The dynamical structure factor is obtained as a sum 
$S=(1+\protect\cos^2\protect\theta)S^{\protect\perp}+
\protect\sin^2\protect\theta S^{||}$, with $\protect\cos\protect\theta=0.1789$
in order to compare with the experimental data. }
\label{fig:s03}
\end{figure}

From the results presented here it seems likely that a contribution to
the structure factor, near $k=0$, from two magnon excitations should be
observable in NENP since $S^\perp(k=0)$ is non zero. Experimentally no
indication of this has been observed so far. Regnault et
al~\cite{regnault92} have performed neutron scattering experiments near
$k=0$ and find no indication of a nonvanishing structure factor
at $k=0$. Ma et al~\cite{ma} analyze their INS data obtained in
the range $k/\pi \geq 0.3$ solely in terms of a single mode and
do not observe the expected two magnon continuum.
These authors point out that the free boson two-magnon prediction for
$S(k,\omega )$,  when convoluted with the experimental
resolution function, yields a much broader peak than what is observed
experimentally, at the smallest value of $k$ studied
experimentally, $k=.3 \pi$, although the integrated intensity is roughly
correct.  In Fig.~\ref{fig:s03} we compare the {\it unconvoluted} free
boson two-magnon prediction at $k=.3\pi$ with the experimental result.  Note
that the integrated intensity and location of the intensity maximum are
approximately correct.
Also note that the width of the theoretical curve is about 2 meV
whereas the experimental width, of about 1 meV, appears to be
resolution limited.  Under these circumstances, the possibility that the
experimentally observed peak is of two-magnon type, with a width
somewhat reduced by interaction effects, cannot be ruled out.  Including
interaction effects in the NL$\sigma$ model does lead to a
narrowing of the two-magnon peak as shown in Ref.~\onlinecite{aw}.
 
The much larger width of the
free boson prediction, of about 5 meV, shown in Ref.~\onlinecite{ma},
Fig. (1c),
is primarily a consequence of convolution with the experimental
resolution
function.  Although this resolution is very narrow in frequency,
it is rather broad in wave-vector,
$\Delta k \approx .25 \pi$.  The relatively narrow convoluted width of
about $1$ meV, was obtained in the experiment by tilting the resolution
ellipsoid so its  axis was approximately parallel to the dispersion
curve.  Under these circumstances, the result of convoluting any
theoretical prediction with the resolution function is extremely
sensitive to the precise shape of the theoretical dispersion relation
(ie. the peak center vs. $k$). It is clear from Figure 18 that the
relativistic two-boson dispersion curve starts to fail badly for $k \geq
.3\pi$.  Hence convoluting this function with a resolution function of
width $.25\pi$ produces an enormous width in frequency.  This is
especially so since the intensity increases rapidly with $k$.
 
Therefore, the very broad resolution of the experiment prevents a
definitive test of the theoretical model.  The extremely good
agreement with the NL$\sigma$ model two-magnon prediction for $S(k)$
shown here, and the shifting of spectral weight above the lowest state,
shown for $L\leq 18$ in Refs.~\onlinecite{golinelli3,tak93}
suggest that the excitations
may start to exhibit a two-particle character at around $k\approx .3\pi$.
Further experimental work with higher resolution and lower $k$ will be
needed to settle this question.

\acknowledgments
We thank 
S.~Ma, C.~Broholm, D.~H.~Reich, B.~J.~Sternlieb, and R.~W.~Erwin
for permission to present some of their data and W.~J.~L.~Buyers for
enlightning discussions. ESS thanks D.~H.~Reich
for several helpful discussion and for suggesting looking at the
single mode approximation for this problem.
ESS also acknowledges helpful discussions with E.~Wong, J.~Gan
and M.~J.~P.~Gingras. This research was supported in part
by NSERC of Canada.

\appendix
\section{Numerical Results}
\begin{table}
\caption{The bulk correlation functions for a
100 site 
$S=1$ antiferromagnetic Heisenberg
chain with single ion anisotropy $D/J=0.18$.}
\label{tab:anisbfuncs}
\begin{tabular}{rdd}
\multicolumn{1}{r}{50-i}
&\multicolumn{1}{d}{$<S^z_{50}S^z_i>$}
&\multicolumn{1}{d}{$<S^x_{50}S^x_i>$}\\
\tableline
0 &	 0.62068642 &	 0.68965678	\\
1 &	-0.40351736 &	-0.49694616	\\
2 &	 0.17899784 &	 0.28905787	\\
3 &	-0.12801833 &	-0.22967616\\
4 &	 0.07923695 &	 0.17667591	\\
5 &	-0.05796982 &	-0.14495068	\\
6 &	 0.03881818 &	 0.11816378	\\
7 &	-0.02870507 &	-0.09867027	\\
8 &	 0.01990824 &	 0.08238886	\\
9 &	-0.01481661 &	-0.06958181	\\
10 &	 0.01048317 &	 0.05886071	\\
11 &	-0.00783740 &	-0.05010433	\\
12 &	 0.00561660 &	 0.04273133	\\
13 &	-0.00421208 &	-0.03658462\\
14 &	 0.00304454 &	 0.03137790	\\
15 &	-0.00228773 &	-0.02698308	\\
16 &	 0.00166324 &	 0.02324034	\\
17 &	-0.00125124 &	-0.02005525	\\
18 &	 0.00091329 &	 0.01733020	\\
19 &	-0.00068747 &	-0.01499761\\
20 &	 0.00050314 &	 0.01299410	\\
21 &	-0.00037883 &	-0.01127166\\
22 &	 0.00027776 &	 0.00978743\\
23 &	-0.00020914 &	-0.00850719	\\
24 &	 0.00015354 &	 0.00740111	\\
25 &	-0.00011557 &	-0.00644516	\\
26 &	 0.00008491 &	 0.00561871	\\
27 &	-0.00006389 &	-0.00489766\\
28 &	 0.00004694 &	 0.00427406	\\
29 &	-0.00003527 &	-0.00373898	\\
30 &	 0.00002586 &	 0.00327238	\\
31 &	-0.00001943 &	-0.00286762	\\
32 &	 0.00001425 &	 0.00251528	\\
33 &	-0.00001070 &	-0.00220953	\\
34 &	 0.00000785 &	 0.00194413\\
35 &	-0.00000589 &	-0.00171480	\\
36 &	 0.00000432 &	 0.00151645	\\
37 &	-0.00000324 &	-0.00134678\\
38 &	 0.00000237 &	 0.00120063	\\
39 &	-0.00000177 &	-0.00107859	\\
40 &	 0.00000129 &	 0.00097339\\
41 &	-0.00000096 &	-0.00089118	\\
42 &	 0.00000070 &	 0.00081782	\\
43 &	-0.00000051 &	-0.00077248\\
44 &	 0.00000036 &	 0.00072126	\\
45 &	-0.00000026 &	-0.00071758\\
46 &	 0.00000018 &	 0.00066999	\\
47 &	-0.00000012 &	-0.00073151	\\
48 &	 0.00000008 &	 0.00062257	\\
49 &	-0.00000006 &	-0.00084534	\\
\end{tabular}
\end{table}

\begin{table}
\caption{The bulk correlation function for a
100 site {\it isotropic}
$S=1$ antiferromagnetic Heisenberg.}
\label{tab:isobfuncs}
\begin{tabular}{rd}
\multicolumn{1}{r}{50-i}
&\multicolumn{1}{d}{$<S^z_{50}S^z_i>$}\\
\tableline
0 &  0.66666666\\
1 & -0.46716133\\
2 &  0.25027190\\
3 & -0.19202160\\
4 &  0.13867476\\
5 & -0.10930536\\
6 &  0.08427910\\
7 & -0.06742565\\
8 &  0.05345277\\
9 & -0.04318563\\
10 &  0.03475723\\
11 & -0.02827864\\
12 &  0.02297530\\
13 & -0.01879232\\
14 &  0.01536715\\
15 & -0.01262184\\
16 &  0.01037002\\
17 & -0.00854571\\
18 &  0.00704591\\
19 & -0.00582170\\
20 &  0.00481286\\
21 & -0.00398488\\
22 &  0.00330106\\
23 & -0.00273757\\
24 &  0.00227127\\
25 & -0.00188587\\
26 &  0.00156641\\
27 & -0.00130176\\
28 &  0.00108208\\
29 & -0.00089974\\
30 &  0.00074818\\
31 & -0.00062220\\
32 &  0.00051740\\
33 & -0.00043016\\
34 &  0.00035753\\
35 & -0.00029698\\
36 &  0.00024654\\
37 & -0.00020437\\
38 &  0.00016925\\
39 & -0.00013976\\
40 &  0.00011524\\
41 & -0.00009450\\
42 &  0.00007736\\
43 & -0.00006271\\
44 &  0.00005074\\
45 & -0.00004051\\
46 &  0.00003212\\
47 & -0.00002578\\
48 &  0.00001909\\
49 & -0.00001897\\
\end{tabular}
\end{table}

\newpage

\end{document}